\documentclass[11pt]{article}
\usepackage{latexsym}
\usepackage{amssymb}
\usepackage{epsf}
\usepackage{amsmath}
\usepackage[hypertex]{hyperref}
\usepackage{graphicx}

\begin{document}
\pagestyle{empty}

\begin{flushright}
LA-UR-08-3715\\
\end{flushright}

\vspace{3cm}

\begin{center}

{\bf\LARGE Study of chargino-neutralino production at hadron
 colliders in a long-lived slepton scenario}
\\

\vspace*{1.5cm}
{\large 
Ryuichiro Kitano
} \\
\vspace*{0.5cm}

{\it Theoretical Division T-8, Los Alamos National Laboratory, Los Alamos, NM 87545}\\
\vspace*{0.5cm}

\end{center}

\vspace*{1.0cm}

\begin{abstract}
{\normalsize
The differential cross section of the chargino-neutralino production, $q
 \bar q \to \chi^\pm \chi^0$, followed by their decays into scalar tau
 leptons, $\chi^\pm \chi^0 \to (\tilde{\tau}^\pm \nu) (\tilde{\tau}^\mp
 \tau^\pm)$ $\to (\tilde{\tau}^\pm \nu) (\tilde{\tau}^\mp l^\pm \nu \bar
 \nu)$, is calculated including the effect of spin correlations. In the
 case where $\tilde{\tau}$ is long-lived, this final state can be fully
 reconstructed in a hadron-collider experiment up to a discrete two-fold
 ambiguity. Distributions of various kinematic variables can thus be
 observable and tell us about masses and spins of superparticles and
 also parity/CP violation in interactions by comparing with the
 cross-section formula. Observing non-trivial distributions derived in
 this paper will be a good test of supersymmetry.
}
\end{abstract} 

\newpage
\baselineskip=18pt
\setcounter{page}{2}
\pagestyle{plain}

\setcounter{footnote}{0}

\section{Introduction}

It is often stated that the LHC is a machine for discovery of new
physics and we will need a new lepton collider to find out what the
actual underlying theory is. This is because most of new physics signals
at the LHC involve multiple jets in final states which are not simple
objects to deal with. It is also true that studies of events with
missing momentum at hadron colliders are more challenging compared to
those at lepton colliders because we cannot use the momentum
conservation in the beam direction. Moreover, unfixed energies of the
initial partons are another obstacle in studying the exclusive
processes.
For this reason, most studies are limited to forming Lorentz (or boost)
invariant quantities out of visible objects to look for peaks, endpoints
or excesses above expected backgrounds. Such kinds of observables do not
usually give enough information to determine the Lagrangian parameters.

Although lepton colliders generally offer a better environment for the
studies of exclusive processes, at hadron colliders it is not impossible
to carry out a detailed study of new-physics events if the final states
are clean enough.
%
In fact, one of the best-motivated models of new physics, supersymmetry
(SUSY), may provide such an opportunity. In the case where the scalar
tau lepton ($\tilde \tau$) is lighter than the neutralinos and
sufficiently long-lived, final states of SUSY events have two charged
tracks of $\tilde \tau$ rather than a missing momentum associated with
escaping neutralinos. The presence of such a long-lived charged particle
significantly improves the capability of the LHC to study SUSY models.

Although the light $\tilde \tau$ scenario has been treated as an
alternative and exotic possibility, it is actually neither theoretically
exotic nor cosmologically problematic.
Since the right-handed $\tilde \tau$ carries only the U(1)$_Y$ quantum
number, quantum corrections to its mass through gauge interactions
are small whereas colored and SU(2) charged sfermions
obtain large positive contributions.
In addition, the Yukawa interaction tends to give a negative
contribution to the mass.
Therefore, it is pretty reasonable to assume that the $\tilde \tau$ is
the lightest among the superpartners of the Standard Model fields.
In such a case, the lifetime of $\tilde \tau$ can be very long although
the estimate depends on the detail of the model; it can decay into a
gravitino and a tau lepton through a suppressed interaction if it is
kinematically allowed or into two Standard Model fermions if $R$-parity
is violated.
There are cosmological constraints on such a long-lived charged
particle~\cite{Gherghetta:1998tq} (\cite{Moroi:1993mb} for related
works), but those can be evaded as long as we do not assume an extremely
long lifetime.
(See \cite{Ibe:2007km} for a recent realistic scenario of supersymmetry
which predicts a long-lived $\tilde \tau$ and naturally explains dark
matter of the Universe by gravitinos.)

There have been studies of the long-lived $\tilde \tau$ at the LHC, and
dramatic differences from the stable neutralino scenario have been
reported. In Ref.~\cite{Hinchliffe:1998ys}, a technique to reconstruct
neutralino masses has been proposed by looking for the decay process
$\chi^0 \to \tilde \tau \tau$. (See \cite{Ibe:2007km, Ellis:2006vu} for
recent studies based on different SUSY models.)
A detailed study of measuring the mass and the momentum of $\tilde \tau$
in the muon system of the ATLAS detector has been done in
Ref.~\cite{Nisati:1997gb,stau,Ambrosanio:2000ik,stauNew}, and it was reported
that the mass can be measured with an accuracy of
$O(0.01-0.1\%)$~\cite{Ambrosanio:2000ik}.
An amusing possibility to collect $\tilde \tau$'s by placing a material
outside the detectors and measure its lifetime has been proposed in
Refs.~\cite{Buchmuller:2004rq, Hamaguchi:2004df, Feng:2004yi}.
Recently, it was pointed out that the spin of $\tilde \tau$ can be
measured by looking at the angular distribution of the pair-production
process of $\tilde \tau$~\cite{Rajaraman:2007ae}.
To discover the long-lived $\tilde \tau$ scenario at hadron colliders,
various signatures have been considered such as highly ionizing
tracks~\cite{Drees:1990yw,Feng:1997zr, Martin:1998vb}, events with
multiple leptons~\cite{Feng:1997zr, Martin:1998vb}, and an excess in the
dimuon-like events~\cite{Feng:1997zr}. (See also
\cite{Dimopoulos:1996yq} for a list of various final states.)
The usefulness of a $p_T$ cut
($p_T$ distribution) in distinguishing a $\tilde \tau$ track from a muon
has been pointed out in Ref.~\cite{Gupta:2007ui}.

In this paper, we study the production process of neutralinos and
charginos followed by their decays into $\tilde \tau$'s. We assume the
lifetime of $\tilde \tau$ is sufficiently long ($\gg$~ns) so that most
of the produced $\tilde \tau$'s reach the muon system where their
three-momentum can be measured. Combined with mass
measurements~\cite{Ambrosanio:2000ik}, one can reconstruct the
four-momentum of the $\tilde \tau$'s.
We mainly focus on the chargino-neutralino production process since it
has the largest cross section among the electroweak production processes
and the final state is rather simple but rich enough to be reconstructed
on an event-by-event basis.
A particularly interesting process is $q \bar q \to \chi^\pm \chi^0 \to
(\tilde \tau^\pm \nu) (\tilde \tau^\mp \tau^\pm) \to (\tilde \tau^\pm
\nu) (\tilde \tau^\mp l^\pm \nu \bar \nu)$, where it is required that
the neutralino decays into $\tilde \tau$ with the opposite charge to the
one from the chargino to avoid a combinatorial background. The
leptonically decaying $\tau$'s are selected so that we can easily
measure the charge of $\tau$. The leptonic mode is also cleaner than
$\tau$-jets with which we need to worry about uncertainties such as fake
jets and the energy scale.
The final state (two opposite-sign $\tilde \tau$'s, a lepton and a
missing momentum) is clean enough to be compared directly with the
theoretical calculation. We present a formula of the cross section
taking into account the spin correlations and demonstrate that various
distributions can be seen at the LHC experiments. These distributions
will be non-trivial tests of SUSY.
Methods to measure the neutralino and chargino masses by using exclusive
processes are also presented.

\section{Interaction Lagrangian}

There are two types of Feynman diagrams for the $\chi^\pm
\chi^0$-production process. One is through an $s$-channel $W$-boson
exchange and the others are the $t$- and $u$-channel squark-exchange
diagrams~\cite{Barger:1983wc}. The interaction Lagrangian for the former
diagram is
\begin{eqnarray}
 {\cal L}_{W} =
\overline{\chi^0} \gamma^\mu
(w_L P_L + w_R P_R ) \chi^- W^+_\mu + {\rm h.c.},
\label{eq:w}
\end{eqnarray}
where $w_L$ and $w_R$ are coupling constants. We will discuss their
relation to the fundamental parameters later. For the squark-exchange
diagrams, the interaction terms are
\begin{eqnarray}
 {\cal L}_{N} =
n_L^{(u)} (\overline{\chi^0} P_L u)
\tilde{u}_L^\dagger 
+ n_L^{(d)} (\overline{\chi^0} P_L d)
\tilde{d}_L^\dagger + {\rm h.c.},
\end{eqnarray}
\begin{eqnarray}
 {\cal L}_{C} =
 c_L^{(u)} (\overline{\chi^+} P_L u)
\tilde{d}_L^\dagger 
+ c_L^{(d)} (\overline{\chi^-} P_L d)
\tilde{u}_L^\dagger + {\rm h.c.},
\label{eq:sq-c}
\end{eqnarray}
where $n_L^{(u,d)}$ and $c_L^{(u,d)}$ are coupling constants.
If we neglect the $u$- and $d$-quark masses, there is no chargino
coupling to the right-handed quarks. Therefore, only the left-handed
(s)quarks participate in the diagrams.

The charginos decay through a term:
\begin{eqnarray}
 {\cal L}_{\chi^-}^D = c_L^{(\nu)} {\overline{\chi^+} P_L \nu_\tau }
  \tilde{\tau}^\dagger + {\rm h.c.}
\end{eqnarray}
There are two terms for the neutralino decay:
\begin{eqnarray}
 {\cal L}_{\chi^0}^D =
\overline{\chi^0} (n_R^{(\tau)} P_R + n_L^{(\tau)} P_L ) \tau
\tilde{\tau}^\dagger + {\rm h.c.}
\label{eq:n}
\end{eqnarray}

\section{The cross-section formula}

\begin{figure}[t]
\begin{center}
\includegraphics[width=8cm]{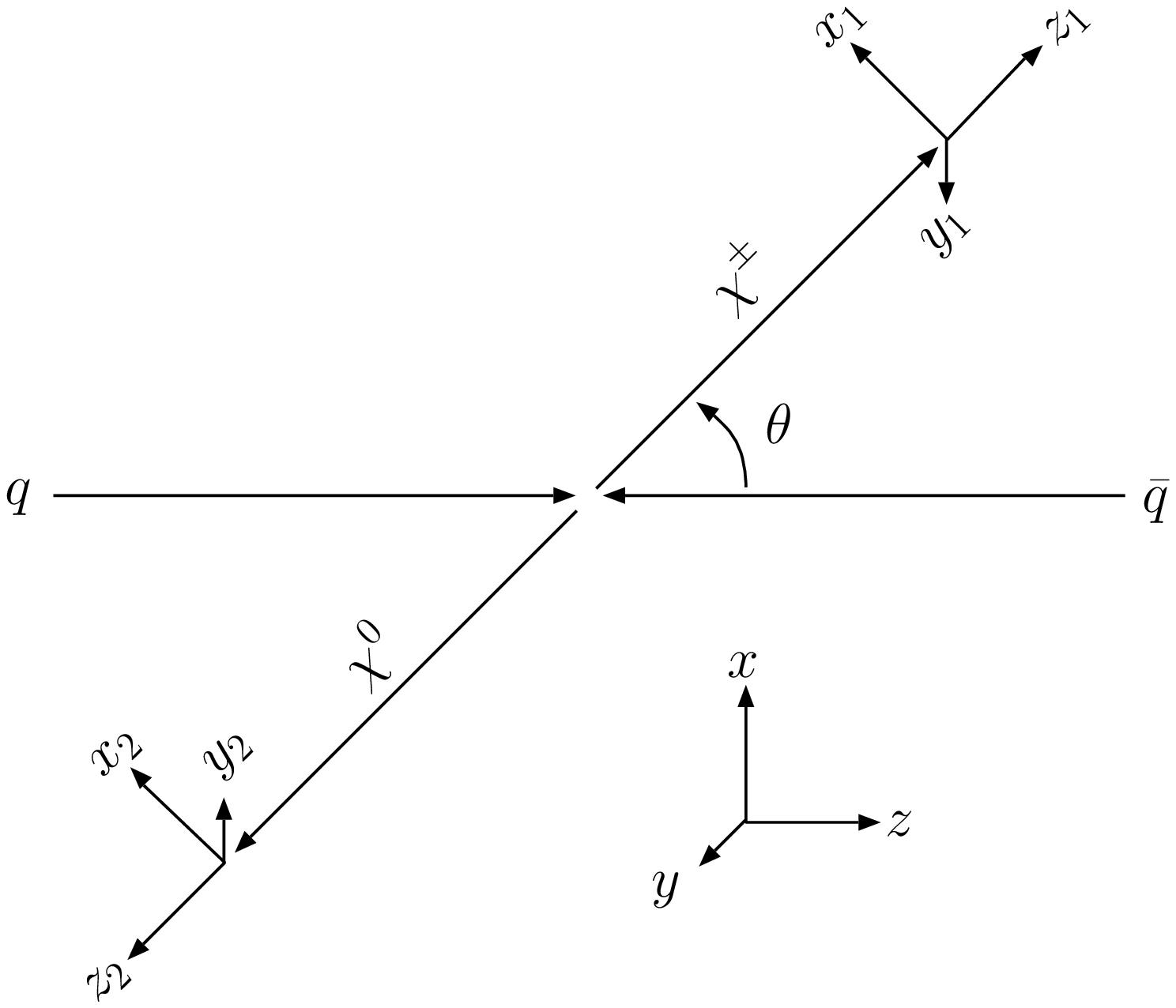}\\
\vspace*{3mm}
\includegraphics[width=10cm]{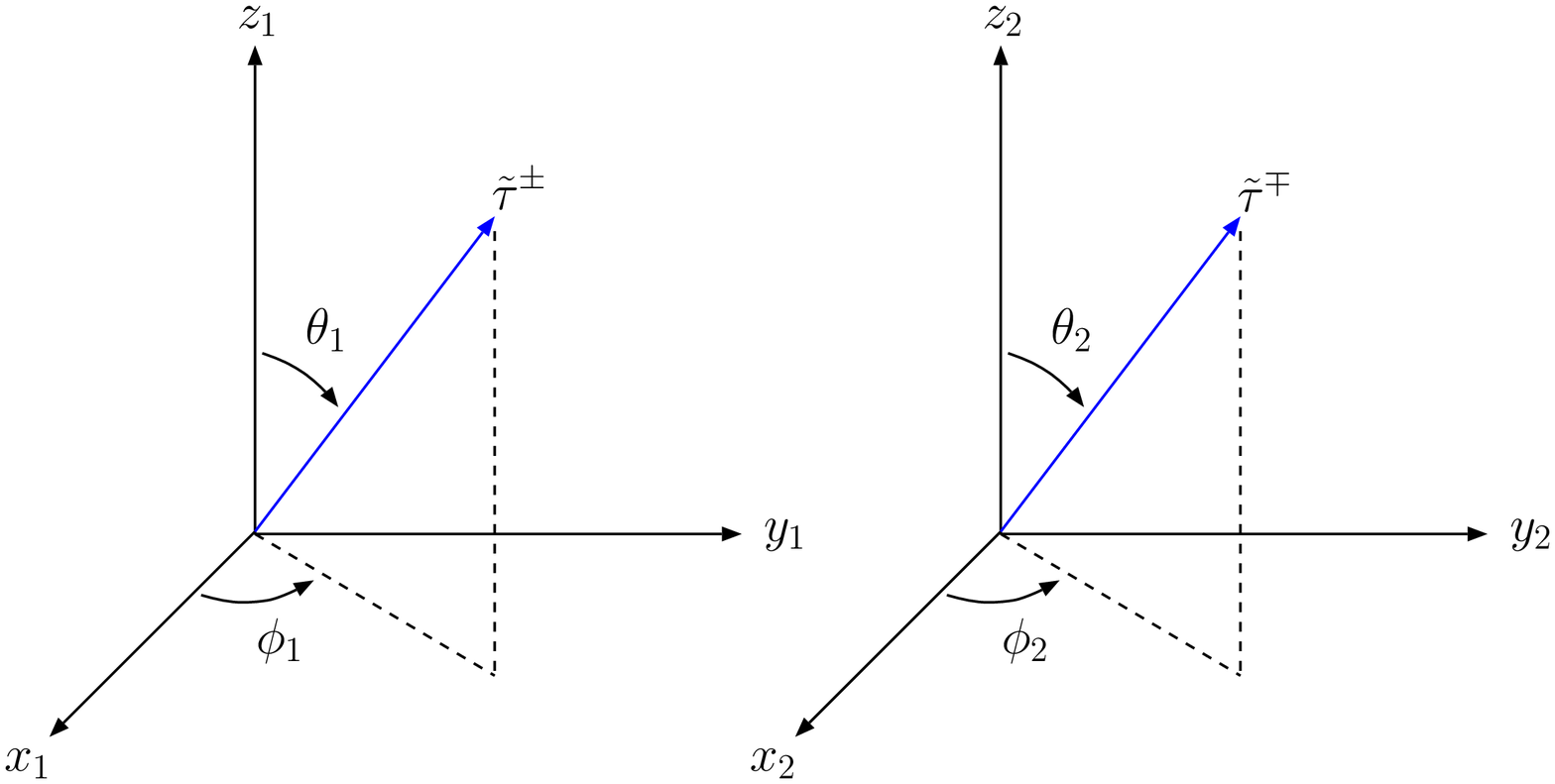} 
\end{center}
\caption{The coordinate systems.}
\label{fig:coordinate}
\end{figure}

We calculate the differential cross section of $q \bar q \to \chi^\pm
\chi^0 \to (\tilde \tau^\pm \nu) (\tilde \tau^\mp \tau^\pm) \to (\tilde
\tau^\pm \nu) (\tilde \tau^\mp l^\pm \nu \bar \nu)$ in terms of the
kinematic variables defined in Fig.~\ref{fig:coordinate}. 
We require the neutralino to decay into $\tilde \tau$ with the opposite
charge to the chargino, and the $\tau$ to decay leptonically.  
From this condition and looking at the charges of the lepton and those
of two $\tilde \tau$'s in the final state, we can tell which $\tilde
\tau$ is from the chargino.
The angle
$\theta$ ($0 \leq \theta \leq \pi $) is defined as the polar angle of
the chargino momentum in the center-of-mass (CM) frame where we take
the production plane to be the $x$-$z$ plane and the direction of the
$x$-axis is chosen so that the $x$-component of the chargino momentum is
positive.
The direction of the $z$-axis is taken to be that of the $q$ momentum.
We also introduce angles $\theta_1$ and $\phi_1$ ($\theta_2$ and
$\phi_2$) which are polar coordinates of the $\tilde \tau^\pm$ ($\tilde
\tau^\mp$) momentum in the rest frame of $\chi^\pm$ ($\chi^0$) ($0 \leq
\theta_{1,2} \leq \pi$ and $0 \leq \phi_{1,2} \leq 2 \pi$).
Momenta in those frames are related by the following Lorentz
transformations:
\begin{eqnarray}
 p_{\rm CM}^\mu &=& \left(
\begin{array}{cccc}
 1 & 0 & 0 & 0 \\
 0 & \cos \theta & 0 & \sin \theta \\
 0 & 0 & 1 & 0 \\
 0 & -\sin \theta & 0 & \cos \theta \\
\end{array}
\right)
\left(
\begin{array}{cccc}
 \gamma_A & 0 & 0 & \gamma_A \beta_A \\
 0 & 1 & 0 & 0 \\
 0 & 0 & 1 & 0 \\
 \gamma_A \beta_A & 0 & 0 & \gamma_A \\
\end{array}
\right) p_1^\mu
\\
&=&
\left(
\begin{array}{cccc}
 1 & 0 & 0 & 0\\
 0 & 1 & 0 & 0 \\
 0 & 0 & -1 & 0 \\
 0 & 0 & 0 & -1 \\
\end{array}
\right)
 \left(
\begin{array}{cccc}
 1 & 0 & 0 & 0 \\
 0 & \cos \theta & 0 & - \sin \theta \\
 0 & 0 & 1 & 0 \\
 0 & \sin \theta & 0 & \cos \theta \\
\end{array}
\right)
\left(
\begin{array}{cccc}
 \gamma_B & 0 & 0 & \gamma_B \beta_B \\
 0 & 1 & 0 & 0 \\
 0 & 0 & 1 & 0 \\
 \gamma_B \beta_B & 0 & 0 & \gamma_B \\
\end{array}
\right) p_2^\mu.
\end{eqnarray}
The boost factors are defined by
\begin{eqnarray}
 \gamma_A =  { 1 + x_A^2 - x_B^2 \over 2 x_A}, \ \ \ 
 \beta_A = \sqrt{ 1 - {1 \over \gamma_A^2} },
\label{eq:gammaA}
\end{eqnarray}
\begin{eqnarray}
 \gamma_B =  { 1 - x_A^2 + x_B^2 \over 2 x_B}, \ \ \
 \beta_B = \sqrt{ 1 - {1 \over \gamma_B^2} },
\label{eq:gammaB}
\end{eqnarray}
where $x_A = m_{\chi^+} / \sqrt{\hat{s}}$ and $x_B = m_{\chi^0} / \sqrt{
\hat{s}}$ with $\hat s = (P_{\chi^+} + P_{\chi^0})^2$.
For later use, we define
\begin{eqnarray}
z_A \equiv 2 x_A \gamma_A,\ \ \ z_B
\equiv 2 x_B \gamma_B,
\label{eq:zAzB}
\end{eqnarray}
which are energies of the chargino and the
neutralino in the CM frame normalized by $\sqrt{\hat{s}} /2$.
Finally, we define 
\begin{eqnarray}
 z_l \equiv {E_l \over E_\tau},\ \ \ (0 \leq z_l \leq 1),
\end{eqnarray}
where the energies of the lepton ($E_l$) and $\tau$ ($E_\tau$) can be
measured in any frame in the approximation $m_\tau \ll m_{\chi^0}$.  In
this limit, the lepton momentum is pointing in the same direction to
that of the parent $\tau$.

The cross-section formula can be written in terms of a product of
density matrices of the production part $\rho^{ab}$ and the decay parts
$D_A^a$ (chargino) and ${\widetilde D}_B^b$ (neutralino)
($a,b=0,...,3$). (See \cite{Haber:1994pe} for example for methods to
calculate the cross section.) By using the narrow width approximation,
it is given by
\begin{eqnarray}
d \sigma \!\!\!&=& \!\!\!
{d \cos \theta \over 2}
{d \Omega_1 \over 4 \pi}
{d \Omega_2 \over 4 \pi}
dz_l
\cdot
{1 \over N_c}
{1 \over 16 \pi}  {g_2^2 \over 2} 
{z_A \beta_A \over \hat{s}}
\left( {1 \over 1 - x_W^2 }\right)^2 
\nonumber \\
&& \times
B(\chi^\pm \to \tilde{\tau}^\pm \nu) 
B(\chi^0 \to \tilde{\tau}^\mp \tau^\pm) 
B(\tau^\pm \to l^\pm \nu \bar \nu) 
\nonumber \\
&& \times
\sum_{a,b=0}^3
D_A^a (\theta_1, \phi_1) \rho^{ab} (\theta) 
{\widetilde D}_B^b (\theta_2, \phi_2, z_l),
\label{eq:cross-section}
\end{eqnarray}
with
\begin{eqnarray}
{\widetilde D}_B^b (\theta_2, \phi_2, z_l) = 
{1 \over 3} ( 1 - z_l )
\left[
(5+ 5z_l - 4 z_l^2) D_B^b (\theta_2, \phi_2)
- a_N ( 1 + z_l - 8 z_l^2) \delta^{b0}
\right],
\label{eq:cross-section2}
\end{eqnarray}
where $g_2$ is the coupling constant of the SU(2)$_L$ gauge interaction,
and $x_W = m_W / \sqrt{\hat{s}}$ with $m_W$ equal to the $W$-boson mass.
The delta factor is simply $\delta^{b0} = (1,0,0,0)$.  A real-number
parameter $a_N$ ($-1 \leq a_N \leq 1$) represents parity violation in
the $\chi^0$-$\tilde \tau$-$\tau$ interaction:
\begin{eqnarray}
 a_N \equiv { | n_L^{(\tau)} |^2 - | n_R^{(\tau)} |^2 
\over | n_L^{(\tau)} |^2 + | n_R^{(\tau)} |^2 }.
\end{eqnarray}
Once we integrate over the lepton-energy fraction, $z_l$, the $D_A \cdot
\rho \cdot {\widetilde D}_B$ part reduces to
\begin{eqnarray}
 D_A \cdot
\rho \cdot {\widetilde D}_B
\to
 D_A \cdot
\rho \cdot D_B.
\end{eqnarray}
Note that the term which is proportional to $a_N$ in
Eq.~(\ref{eq:cross-section2}) vanishes after the integration over $z_l$.

The decay parts $D^a_A$ and $D^b_B$ have a simple form:
\begin{eqnarray}
 D^a_A = \left(
\begin{array}{c}
 1\\
\pm a_C \sin \theta_1 \cos \phi_1 \\
\pm a_C \sin \theta_1 \sin \phi_1 \\
\pm a_C \cos \theta_1\\
\end{array}
\right), \ \ \ 
 D^b_B = \left(
\begin{array}{c}
 1\\
\mp a_N  \sin \theta_2 \cos \phi_2 \\
\mp a_N  \sin \theta_2 \sin \phi_2 \\
\mp a_N  \cos \theta_2\\
\end{array}
\right),
\end{eqnarray}
where $a_C$ is the parity-violation factor in the chargino decay. It
always takes the maximum value:
\begin{eqnarray}
 a_C = 1,
\end{eqnarray}
due to the fact that the neutrinos have only the left-handed
chirality.
Each component of $D_A$ and $D_B$ corresponds to the expansion
coefficient of the Hermitian $2 \times 2$ spin-density matrices in terms
of the unit ($a,b=0$) and the Pauli ($a,b=1,...,3$) matrices. The
non-trivial dependencies on angles appear if there is parity violation
in the decay vertices.
An integration over a solid angle $d \Omega_1$ ($d\Omega_2$) leads to
\begin{eqnarray}
 D_A \cdot \rho \cdot \widetilde D_B \to
 \rho^{0b} \cdot \widetilde D_B,\ \ \ 
( D_A \cdot \rho \cdot \widetilde D_B \to
 D_A \cdot \rho^{a0} \widetilde D_B^0).
\end{eqnarray}
When we perform a further integration of angles, $d\Omega_1 d\Omega_2$,
and $z_l$, we obtain
\begin{eqnarray}
 D_A \cdot \rho \cdot \widetilde D_B \to
 \rho^{00} \widetilde D_B^0 \to \rho^{00}.
\end{eqnarray}

The production part $\rho^{ab}$ is expressed in terms of $\hat s$, the angle 
$\theta$ and effective coupling factors $\bar{w}_L$ and $\bar{w}_R$ defined by
\begin{eqnarray}
 \bar{w}_L \equiv w_L 
- {1 \over 2}
{ 1 - x_W^2 \over x_{\tilde{u}_L}^2 - \hat t/\hat{s} }
{ c_L^{(u)} n_L^{(d)*} \over g_2 / \sqrt{2}},
\end{eqnarray}
\begin{eqnarray}
 \bar{w}_R \equiv w_R 
+ {1 \over 2}
{ 1 - x_W^2 \over x_{\tilde{d}_L}^2 - \hat u/\hat{s} }
{ c_L^{(d)*} n_L^{(u)} \over g_2 / \sqrt{2}},
\end{eqnarray}
where
\begin{eqnarray}
 \hat t = - \hat{s} z_A  \cdot { 1 \mp \beta_A \cos \theta  \over 2 },\ \ \ 
 \hat u = - \hat{s} z_B  \cdot { 1 \pm \beta_B \cos \theta  \over 2 },
\end{eqnarray}
and
\begin{eqnarray}
 x_{\tilde{u}_L} = { m_{\tilde{u}_L} \over \sqrt{ \hat{s} } }, \ \ \ 
 x_{\tilde{d}_L} = { m_{\tilde{d}_L} \over \sqrt{ \hat{s} }}.
\end{eqnarray}
The masses $m_{\tilde{u}_L}$ and $m_{\tilde{d}_L}$ are those of the
left-handed squarks, $\tilde u_L$ and $\tilde d_L$, respectively.
The components $\rho^{ab}$ are given by
\begin{eqnarray}
 \rho^{00} &=& 
{1 \over 4 }(|\bar{w}_L|^2 + | \bar{w}_R |^2 ) 
z_A z_B ( 1 + \beta_A \beta_B \cos^2 \theta)
\nonumber \\
&& + 2 {\rm Re} (\bar{w}_L^* \bar{w}_R) x_A x_B  
\nonumber \\
&& \mp {1 \over 2 }(|\bar{w}_L|^2 - | \bar{w}_R |^2 ) 
z_A \beta_A \cos \theta,
\end{eqnarray}
\begin{eqnarray}
 \rho^{01} &=& 
{1 \over 2}(|\bar{w}_L|^2 + | \bar{w}_R |^2 ) z_A x_B \sin \theta
\nonumber \\
&& + {\rm Re} [ \bar{w}_L^* \bar{w}_R ] 
x_A z_B \sin \theta
\nonumber \\
&& \mp {1 \over 2}(|\bar{w}_L|^2 - | \bar{w}_R |^2 )
z_A x_B \beta_A \cos \theta \sin \theta,
\end{eqnarray}
\begin{eqnarray}
 \rho^{02} &=& 
{\rm Im} [ \bar{w}_L^* \bar{w}_R ] 
x_A z_B \beta_B \sin \theta,
\end{eqnarray}
\begin{eqnarray}
 \rho^{03} &=& 
{1\over 4}
(|\bar{w}_L|^2 + | \bar{w}_R |^2 ) z_A z_B (1 + \beta_A \beta_B ) \cos \theta
\nonumber \\
&& + 2 {\rm Re} [ \bar{w}_L^* \bar{w}_R ] x_A x_B \cos \theta
\nonumber \\
&& \mp {1 \over 4}
(|\bar{w}_L|^2 - | \bar{w}_R |^2 ) z_A z_B
( \beta_B + \beta_A \cos^2 \theta ),
\end{eqnarray}
\begin{eqnarray}
 \rho^{10} &=&
{1 \over 2}
(|\bar{w}_L|^2 + | \bar{w}_R |^2 ) x_A z_B \sin \theta
\nonumber \\
&& + {\rm Re} [ \bar{w}_L^* \bar{w}_R ] z_A x_B \sin \theta
\nonumber \\
&& \mp {1\over2}
(|\bar{w}_L|^2 - | \bar{w}_R |^2 ) x_A z_A \beta_A \cos \theta
 \sin \theta,
\end{eqnarray}
\begin{eqnarray}
 \rho^{11} &=& 
(|\bar{w}_L|^2 + | \bar{w}_R |^2 ) x_A x_B \sin^2 \theta
\nonumber \\
&& + {1 \over 2} {\rm Re} [ \bar{w}_L^* \bar{w}_R ] z_A z_B \sin^2 \theta,
\end{eqnarray}
\begin{eqnarray}
 \rho^{12} &=& 
- {1\over2} {\rm Im} [ \bar{w}_L^* \bar{w}_R ] 
z_A z_B \beta_B \sin^2 \theta,
\end{eqnarray}
\begin{eqnarray}
 \rho^{13} &=& 
{1\over2}
(|\bar{w}_L|^2 + | \bar{w}_R |^2 ) x_A z_B \cos \theta \sin \theta
\nonumber \\
&& + {\rm Re} [ \bar{w}_L^* \bar{w}_R ] z_A x_B \cos \theta \sin \theta
\nonumber \\
&& \mp {1\over2}(|\bar{w}_L|^2 - | \bar{w}_R |^2 ) x_A z_B \beta_B \sin \theta,
\end{eqnarray}
\begin{eqnarray}
 \rho^{20} &=&
{\rm Im} [ \bar{w}_L^* \bar{w}_R ] 
x_B z_A \beta_A \sin \theta,
\end{eqnarray}
\begin{eqnarray}
 \rho^{21} &=& 
{1 \over 2} {\rm Im} [ \bar{w}_L^* \bar{w}_R ] 
z_A z_B \beta_A \sin^2 \theta,
\end{eqnarray}
\begin{eqnarray}
 \rho^{22} &=&
{1 \over 2} {\rm Re} [ \bar{w}_L^* \bar{w}_R ]
z_A z_B \beta_A \beta_B \sin^2 \theta,
\end{eqnarray}
\begin{eqnarray}
 \rho^{23} &=& 
- {\rm Im} [ \bar{w}_L^* \bar{w}_R ] 
z_A x_B \beta_A \cos \theta \sin \theta,
\end{eqnarray}
\begin{eqnarray}
 \rho^{30} &=& 
- {1 \over 4}
(|\bar{w}_L|^2 + | \bar{w}_R |^2 ) z_A z_B (1 + \beta_A\beta_B ) \cos \theta
\nonumber \\
&& - 2 {\rm Re} [ \bar{w}_L^* \bar{w}_R ] x_A x_B \cos \theta
\nonumber \\
&& \pm {1 \over 4}(|\bar{w}_L|^2 - | \bar{w}_R |^2 ) z_A z_B
( \beta_A + \beta_B \cos^2 \theta ),
\end{eqnarray}
\begin{eqnarray}
 \rho^{31} &=& 
- {1 \over 2}
(|\bar{w}_L|^2 + | \bar{w}_R |^2 ) z_A x_B \cos \theta \sin \theta
\nonumber \\
&& - {\rm Re} [ \bar{w}_L^* \bar{w}_R ] 
x_A z_B \cos \theta \sin \theta
\nonumber \\
&& \pm {1\over2}
(|\bar{w}_L|^2 - | \bar{w}_R |^2 ) x_B z_A \beta_A \sin \theta,
\end{eqnarray}
\begin{eqnarray}
 \rho^{32} &=& 
{\rm Im} [ \bar{w}_L^* \bar{w}_R ] 
z_B x_A \beta_B \cos \theta \sin \theta,
\end{eqnarray}
\begin{eqnarray}
 \rho^{33} &=& 
- {1\over4}
(|\bar{w}_L|^2 + | \bar{w}_R |^2 )
z_A z_B ( \beta_A \beta_B + \cos^2 \theta)
\nonumber \\
&& - 2 {\rm Re} [ \bar{w}_L^* \bar{w}_R ] 
x_A x_B \cos^2 \theta
\nonumber \\
&& \pm {1\over2}(|\bar{w}_L|^2 - | \bar{w}_R |^2 ) 
z_A \beta_A \cos \theta.
\end{eqnarray}
With a fixed $\hat s$, the energies $z_A (\equiv
2E_{\chi^\pm}/\sqrt{\hat s})$ and $z_B (\equiv 2E_{\chi^0}/\sqrt{\hat
s})$ and the velocities $\beta_A$ and $\beta_B$ are constants as defined
in Eqs.~(\ref{eq:gammaA}), (\ref{eq:gammaB}) and (\ref{eq:zAzB}). The
spin summed part $\rho^{00}$ has also been calculated in
Ref.~\cite{Barger:1983wc}.

By using this cross-section formula we will be able to extract various
information such as parity and CP violating parameters in the
interaction Lagrangian.

\section{Asymmetries vs parameters in the Lagrangian}

The cross-section formula derived in the previous section further
simplifies in the case where the left-handed squarks are much heavier
than the $\chi^0/\chi^\pm$ in the intermediate state, or one of the
$\chi^0/\chi^\pm$ is Higgsino-like. In such cases, the diagrams with
squark exchanges are not important and the angular dependencies in $\bar
w_L$ and $\bar w_R$ vanish. In this approximation,
\begin{eqnarray}
 \bar w_L = w_L,\ \ \ \bar w_R = w_R.
\end{eqnarray}
This situation is not unrealistic since quantum corrections tend to make
the squarks much heavier than other superparticles. In the following
discussion we will use this simplification. For a more general analysis,
one should use the full formula derived in the previous section.

\begin{figure}[t]
\begin{center}
\includegraphics[width=7.6cm]{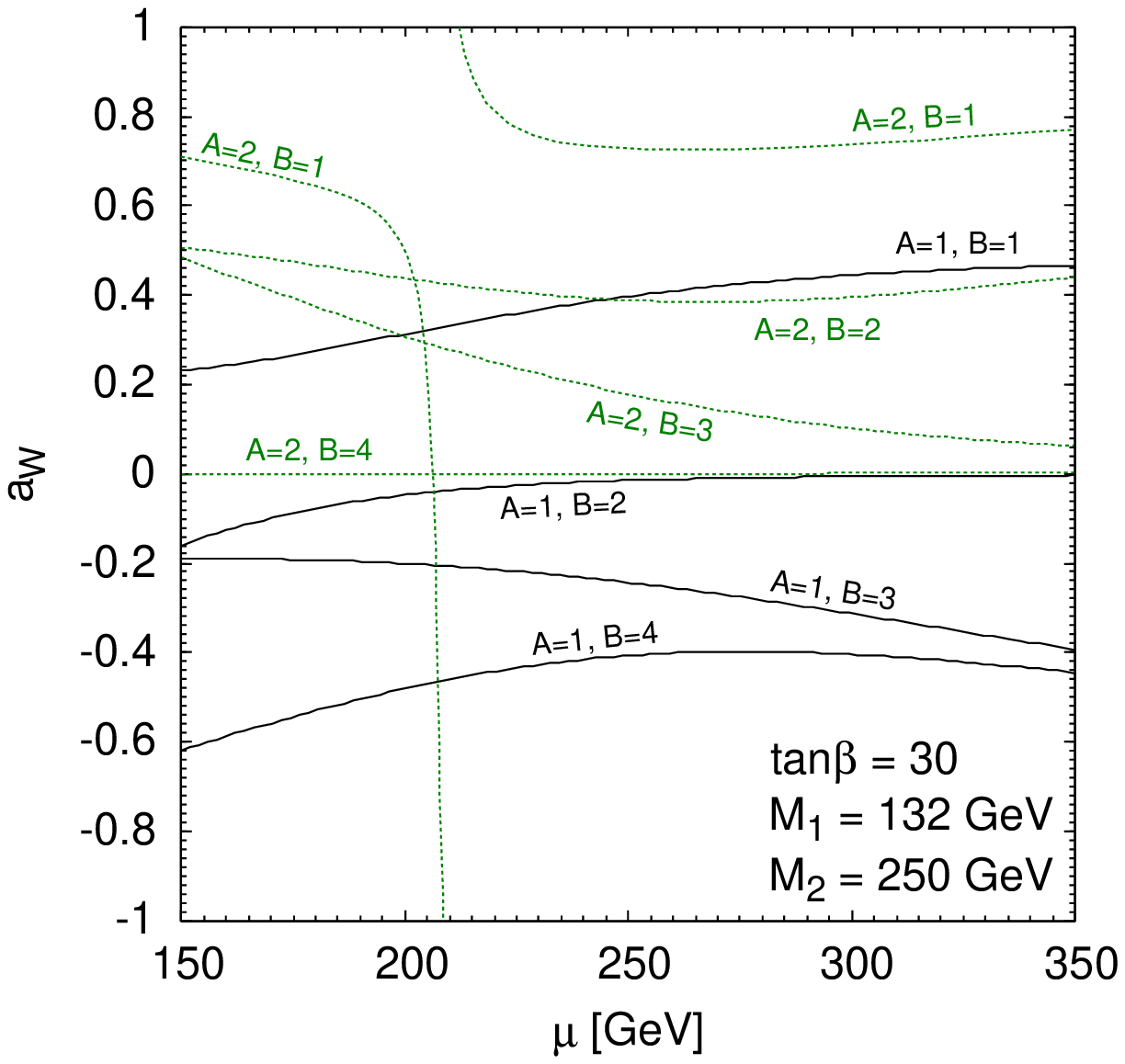}
\includegraphics[width=7.5cm]{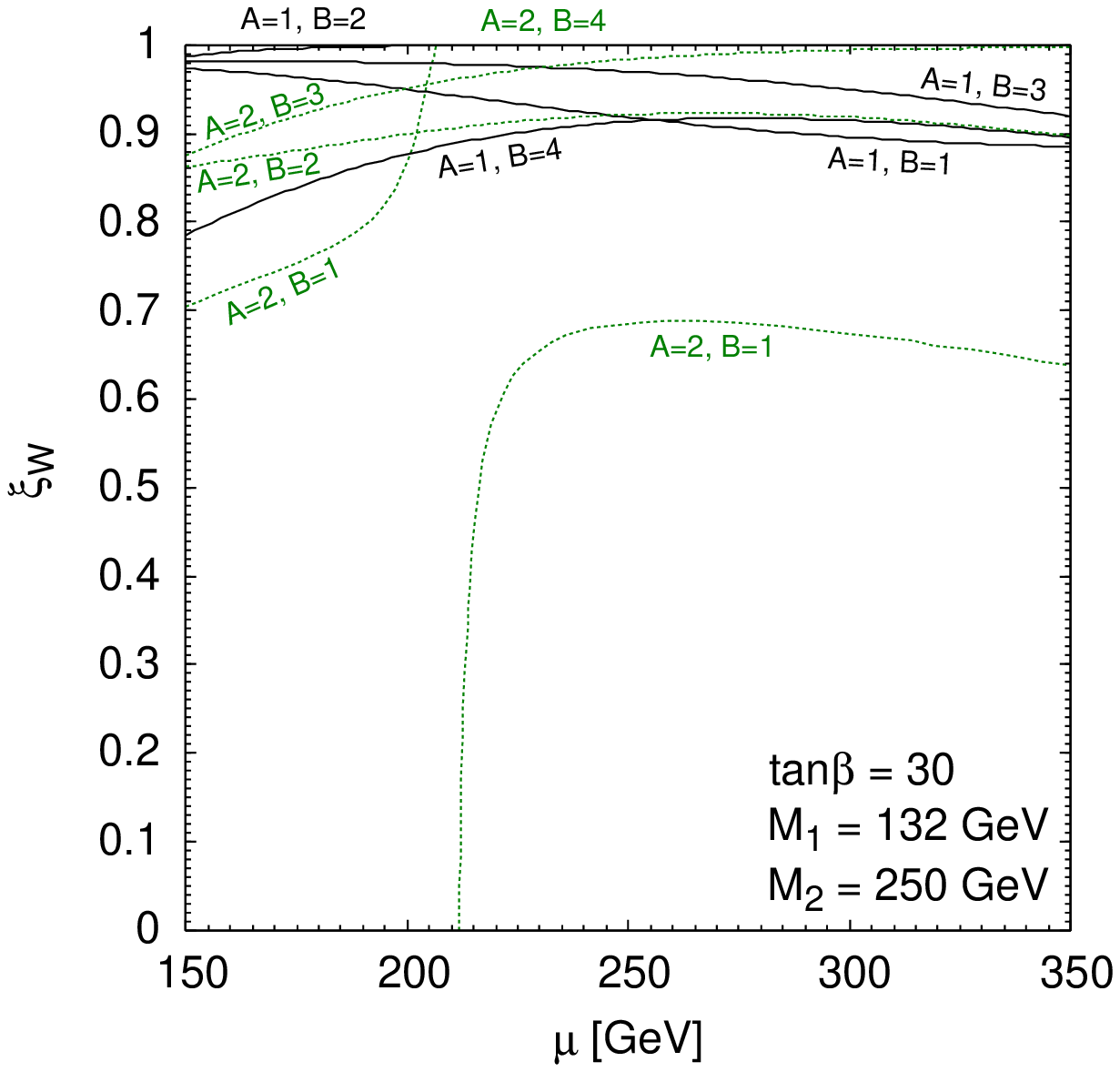}
\end{center}
\caption{The $\mu$ parameter dependence of $a_W$ and $\xi_W$. The labels
 $A = 1,2$ and $B=1-4$ represent each mass eigenstate of the charginos
 and the neutralinos.}  \label{fig:aW}
\end{figure}

\begin{figure}[t]
\begin{center}
 \includegraphics[width=7.5cm]{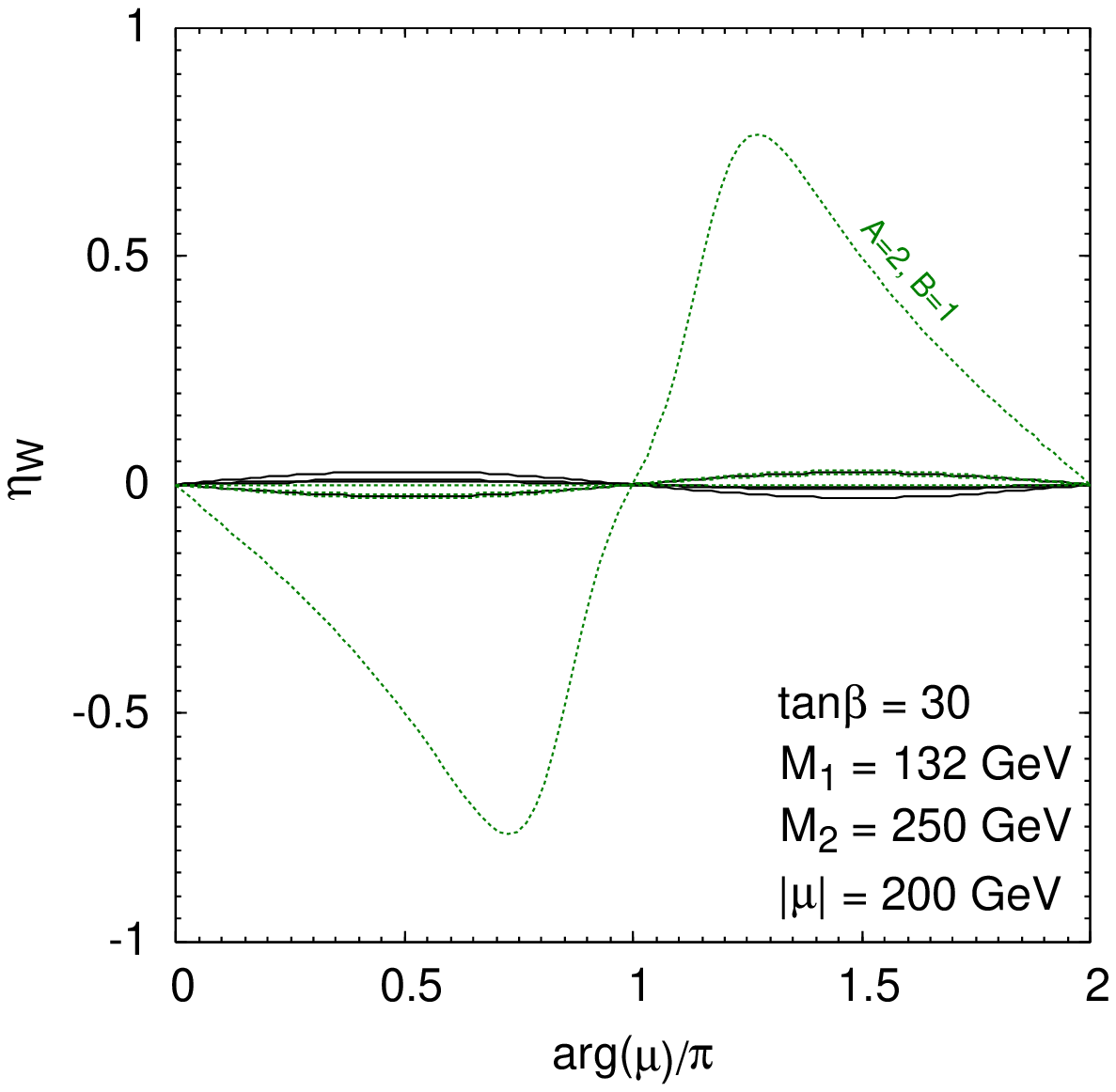}
 \includegraphics[width=7.5cm]{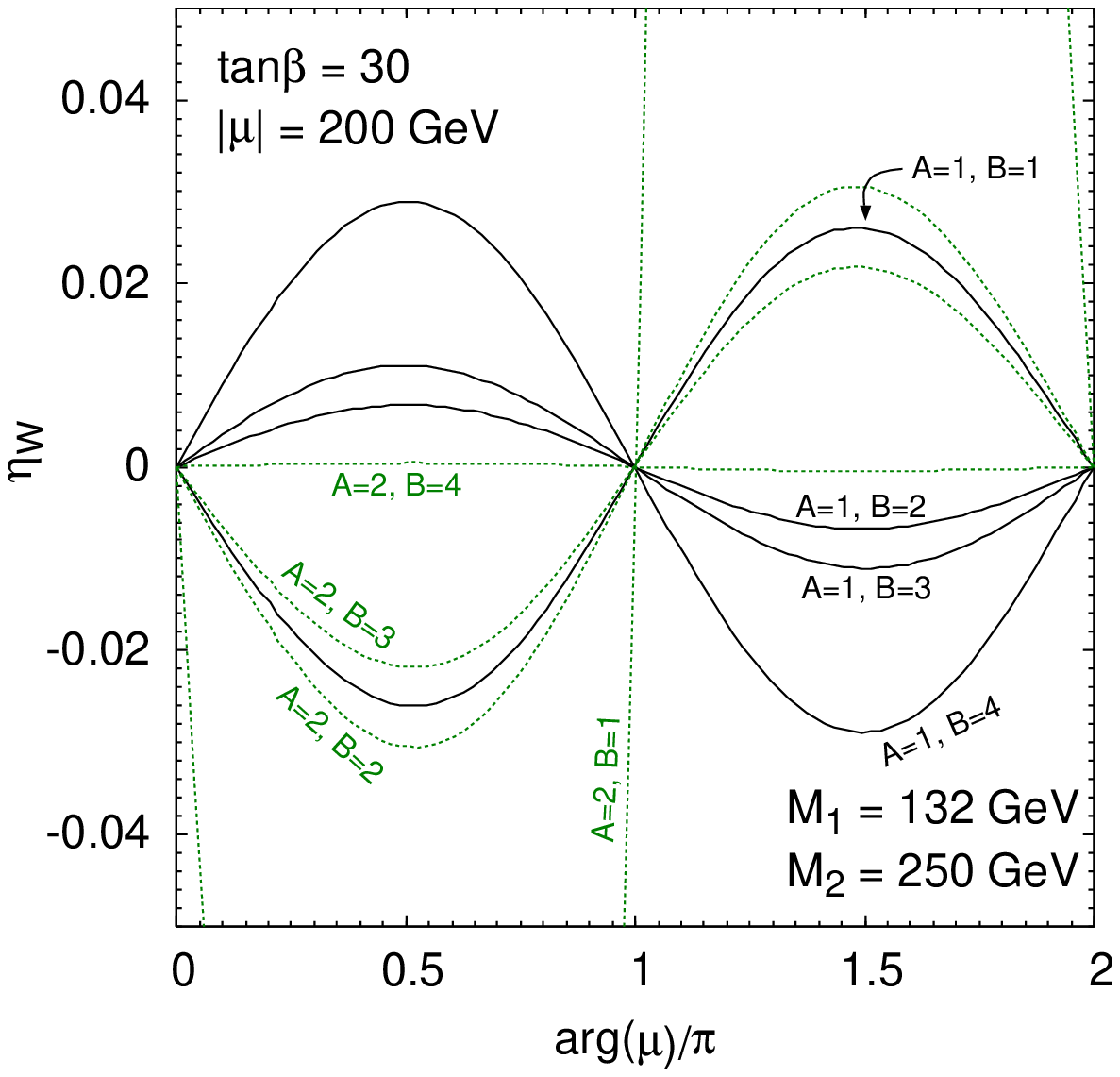}
\end{center}
\caption{The phase dependence of the CP asymmetry $\eta_W$. The right
 figure is the same as the left figure with a different scale.}
\label{fig:etaW}
\end{figure}

We have defined two parity-asymmetry parameters $a_N$ and $a_C (\equiv
1)$ for the decay processes in the previous section. For the production part
$\rho$, we define the following three quantities:
\begin{eqnarray}
 a_W = {|w_L|^2 - |w_R|^2 \over |w_L|^2 + |w_R|^2},\ \ \ 
 \xi_W = {2 {\rm Re}[w_L^* w_R] \over |w_L|^2 + |w_R|^2},\ \ \ 
 \eta_W = {2 {\rm Im}[w_L^* w_R] \over |w_L|^2 + |w_R|^2}.
\end{eqnarray}
The matrix $\rho$ can be expressed in terms of the three quantities and
an angle $\theta$.
In this section, we discuss model parameters and their relations to the
observables ($a_N$, $a_W$, $\xi_W$, $\eta_W$) defined here.

In the minimal supersymmetric standard model, there are five model
parameters which are relevant for the process: the Higgsino mass
parameter $\mu$, the ratio of the vacuum expectation values of the Higgs
fields $\tan \beta (\equiv \langle H_2 \rangle / \langle H_1 \rangle )$,
the gaugino mass parameters $M_1$ and $M_2$, and the mixing parameter of
the scalar tau leptons $\theta_{\tilde \tau}$ ($\tilde \tau_1 = \cos
\theta_{\tilde \tau} \tilde \tau_R + \sin \theta_{\tilde \tau} \tilde
\tau_L$).

The coupling constants $w_{L,R}$ and $n^{(\tau)}_{L,R}$ in
Eqs.~(\ref{eq:w}) and (\ref{eq:n}) are expressed in terms of mixing
matrices of the neutralinos $O_N$, of the charginos $O_L$ and $O_R$, and
the mixing angle of the scalar tau leptons $\theta_{\tilde \tau}$. The
matrices are defined by
\begin{eqnarray}
 O_N M_{\chi^0} O_N^{T} = M_{\chi^0}^{\rm diag.},
\end{eqnarray}
\begin{eqnarray}
 O_R M_{\chi^+} O_L^\dagger = M_{\chi^+}^{\rm diag.},
\end{eqnarray}
where the right-hand-side of the equations are diagonal matrices with
real and positive eigenvalues. The mass matrices $M_{\chi^0}$ and
$M_{\chi^+}$ are 
\begin{eqnarray}
 M_{\chi^0} = \left(
\begin{array}{cccc}
 M_1 & 0  & \displaystyle {g_Y v \over \sqrt 2} \cos \beta 
& \displaystyle - {g_Y v \over \sqrt 2} \sin \beta \\ 
 0 & M_2  & \displaystyle - {g_2 v \over \sqrt 2} \cos \beta 
& \displaystyle {g_2 v \over \sqrt 2} \sin \beta \\ 
 \displaystyle {g_Y v \over \sqrt 2} \cos \beta & 
 \displaystyle - {g_2 v \over \sqrt 2} \cos \beta &
0 & - \mu \\ 
 \displaystyle - {g_Y v \over \sqrt 2} \sin \beta & 
 \displaystyle  {g_2 v \over \sqrt 2} \sin \beta &
- \mu & 0 \\ 
\end{array}
\right),
\end{eqnarray}
and
\begin{eqnarray}
 M_{\chi^-} = \left(
\begin{array}{cc}
 M_2 & -g_2 v \cos \beta \\
 -g_2 v \sin \beta & \mu \\
\end{array}
\right).
\end{eqnarray}
The vacuum expectation value $v$ is $v = \sqrt{\langle H_1 \rangle^2 + \langle
H_2 \rangle^2} = 174$~GeV.

In terms of the mixing matrices, the coupling constants are given by
\begin{eqnarray}
 w_L &=& g_2 (O_N^*)_{B2} (O_L^*)_{A1} 
+ {g_2 \over \sqrt 2} (O_N^*)_{B3} (O_L^*)_{A2}, \label{eq:wL}\\
 w_R &=& g_2 (O_N)_{B2} (O_R^*)_{A1} 
- {g_2 \over \sqrt 2} (O_N)_{B4} (O_R^*)_{A2},
\label{eq:wR}
\end{eqnarray}
and
\begin{eqnarray}
 n_L^{(\tau)} &=&
- {g_2 \over \sqrt 2} (O_N)_{B2} \sin \theta_{\tilde \tau}
- {g_Y \over \sqrt 2} (O_N)_{B1} \sin \theta_{\tilde \tau}
- {m_\tau \over v \cos \beta} (O_N)_{B3} \cos \theta_{\tilde \tau},\\
 n_R^{(\tau)} &=&
\sqrt 2 g_Y (O_N^*)_{B1} \cos \theta_{\tilde \tau}
- {m_\tau \over v \cos \beta} (O_N^*)_{B3} \sin \theta_{\tilde \tau},
\end{eqnarray}
where $g_Y$ is the coupling constant of the U(1)$_Y$ gauge interaction.
The subscripts for the mixing matrices indicate their corresponding
components. The indices $A(=1,2)$ and $B(=1,...,4)$ represent mass
eigenstates of the charginos and the neutralinos, respectively. The
second indices of $O_L$, $O_R$ and $O_N$ are the ones for the
interaction eigenbasis; (Wino, Higgsino) for charginos and (Bino, Wino,
down-type Higgsino, up-type Higgsino) for neutralinos.

We show in Fig.~\ref{fig:aW} the $a_W$ and $\xi_W$ factors in a limited
case where we fix ($\tan \beta$, $M_1$, $M_2$) to be (30, 132~GeV,
250~GeV) and vary the $\mu$ parameter from 150~GeV to 350~GeV.  The
labels $A = 1,2$ and $B=1,...,4$ represent each mass eigenstate of the
charginos and the neutralinos, respectively.  
For a small value of $\mu$, the lighter chargino ($A=1$) is
Higgsino-like. As $\mu$ increases, the lighter chargino goes through the
mixed region ($\mu \sim M_2$) to the Wino-like region ($\mu \gg M_2$).
The largest cross section is for $A=1$ and $B=2$ which gives $a_W \simeq
0$ and $\xi_W \simeq 1$.
In the two extreme limits where the produced chargino and neutralino are
both purely Higgsinos or both purely Winos, there is no parity violation
in the interaction vertex (i.e., $a_W = 0$) since they are vector-like
particles. A large deviation from $a_W \sim 0$ is possible when there is
a significant mixing among the Higgsinos, Wino and Bino.

The CP asymmetry $\eta_W$ is calculated with varying the phase of the
$\mu$ parameter in Fig.~\ref{fig:etaW}. The parameters are fixed as
($\tan \beta$, $M_1$, $M_2$, $|\mu|$) $=$ (30, 132~GeV, 250~GeV,
200~GeV). A large CP asymmetry is obtained only for $A=2$ and
$B=1$. Other asymmetries are at most of order a few percent. (Note that
the relative phase between $M_1$ and $M_2$, arg$(M_1 M_2^*)$, is also an
independent physical parameter.)

\begin{figure}[t]
\begin{center}
 \includegraphics[width=7.5cm]{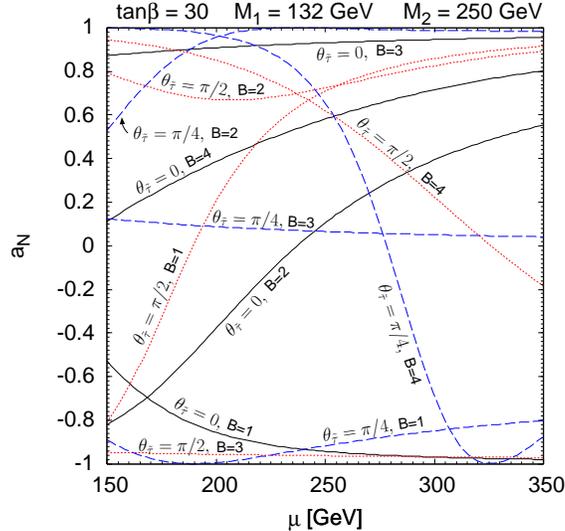}
\end{center}
\caption{The $\mu$ parameter dependence of the parity asymmetry $a_N$.} 
\label{fig:aN}
\end{figure}

Finally, the parity asymmetry $a_N$ in the neutralino decay is
calculated with ($\tan \beta$, $M_1$, $M_2$) $=$ (30, 132~GeV, 250~GeV)
in Fig.~\ref{fig:aN}. We take three values of the $\tilde \tau$-mixing
parameter $\theta_{\tilde \tau} = 0, \pi/4, \pi/2$. Fig.~\ref{fig:aN}
shows that $a_N$ is highly dependent on the model parameters;
specifically the properties of the neutralino and $\tilde \tau$.
This parameter does not necessarily vanish in the pure Higgsino or Wino
limits due to the chiral nature of the tau lepton.
For example, if the neutralino is purely Higgsino and the stau is
left-handed (right-handed), the tau lepton must be right-handed
(left-handed), i.e., $a_N = -1$ (+1), although we need to take into
account mixings in more realistic cases.

We can see that the asymmetry parameters, especially $a_N$, are
sensitive to the model parameters. As we will see in the next section,
the parameters, $a_N$, $a_W$, $\eta_W$ can be measured by looking at
asymmetries in various distributions. Measurements of those asymmetries
together with the mass measurements provide us with information on
combinations of the neutralino/chargino mixings and the $\tilde \tau$
mixing.

\section{Angular and energy distributions}
\label{sec:dist}

We can obtain various one-dimensional distributions by integrating over
the remaining variables in the differential cross section
(\ref{eq:cross-section}). Here we adopt the simplification that the
squark diagrams are not important.


We list in Table~\ref{tab:cpt} transformation properties of the angles
under the charge conjugation (C), the parity transformation (P), and the
time reversal (T).
The transformation properties of asymmetry parameters are assigned in
the right table. With these assignments, the interaction Lagrangian in
Eqs.~(\ref{eq:w}--\ref{eq:n}) are ``formally'' C, P, and T invariant.
These are helpful in understanding the resulting distributions.

\begin{table}[t]
\begin{center}
 \begin{tabular}{|c|c|c|c|}
  \hline
   & C & P & T \\ \hline \hline
  $\cos \theta$ & $-$ & $+$ & $+$ \\
  $\sin \theta$ & $+$ & $+$ & $+$ \\ \hline 
  $\cos \theta_{1,2}$ & $+$ & $+$ & $+$ \\
  $\sin \theta_{1,2}$ & $+$ & $+$ & $+$ \\ \hline
  $\cos \phi_{1,2}$ & $-$ & $+$ & $+$ \\
  $\sin \phi_{1,2}$ & $-$ & $-$ & $-$ \\ \hline
 \end{tabular}
\hspace{2mm}
 \begin{tabular}{|c|c|c|c|}
  \hline
   & C & P & T \\ \hline \hline
  $a_W$ & $-$ & $-$ & $+$ \\
  $\xi_W$ & $+$ & $+$ & $+$ \\
  $\eta_W$ & $+$ & $-$ & $-$ \\
  $a_C$ & $-$ & $-$ & $+$ \\ 
  $a_N$ & $-$ & $-$ & $+$ \\ 
  $\pm$ & $-$ & $+$ & $+$ \\ \hline
 \end{tabular}
\end{center}
\caption{C, P, and T transformation properties of angles defined in
 Fig.~\ref{fig:coordinate} and of the asymmetry parameters. With these
 assignments in the right table the interaction Lagrangian in
 Eqs.~(\ref{eq:w}--\ref{eq:n}) are ``formally'' C, P, and T
 invariant. The symbol $\pm$ represents the charge of the chargino.}
 \label{tab:cpt}
\end{table}

\subsection*{\boldmath{$ z_l $} distribution}

By integrating over $\theta$, ($\theta_1$, $\phi_1$), and ($\theta_2$,
$\phi_2$), we obtain
\begin{eqnarray}
 d\sigma &=& 
\sigma (q \bar q \to \chi^\pm \chi^0)
B(\chi^\pm \to \tilde{\tau}^\pm \nu)
B(\chi^0 \to \tilde{\tau}^\mp \tau^\pm)
B(\tau^\pm \to l^\pm \nu \bar \nu) d z_l
\nonumber \\
&& \times 
{1 \over 3} ( 1 - z_l )
\left[
(5 + 5 z_l - 4 z_l^2 )
- a_N ( 1 + z_l - 8 z_l^2 )
\right].
\label{eq:z_l}
\end{eqnarray}
This is the well-known polarization dependence of the lepton-energy
distribution occurring in leptonic $\tau$ decays~\cite{Bullock:1992yt}.
Since $z_l$ is a rotation and boost invariant quantity (in the limit of
$m_\tau / m_{\chi^0} \ll 1$), we can measure this distribution in the
laboratory frame. This distribution will tell us about the parity
asymmetry $a_N$ in the neutralino decay through the $\tau$ polarization.
This distribution will remain unchanged when we include the squark diagrams.

\subsection*{\boldmath $\cos \theta_1$ distribution}

By integrating over $\theta$, $\phi_1$, ($\theta_2$, $\phi_2$), and
$z_l$, we obtain
\begin{eqnarray}
  d\sigma &=& 
\sigma (q \bar q \to \chi^\pm \chi^0)
B(\chi^\pm \to \tilde{\tau}^\pm \nu)
B(\chi^0 \to \tilde{\tau}^\mp \tau^\pm)
B(\tau^\pm \to l^\pm \nu \bar \nu) {d \cos \theta_1 \over 2}
\nonumber \\
&& \times 
\left[ 1 + a_W a_C f_1(\beta_A, \beta_B) \cos \theta_1 \right].
\label{eq:cos1}
\end{eqnarray}
Recall that $a_C = 1$.
%
The function $f_1$ is given by
\begin{eqnarray}
\displaystyle
 f_1(\beta_A, \beta_B) = 
{
3 \beta_A +  \beta_B  
\over 
3 + \beta_A \beta_B + 3 \xi_W \sqrt{(1- \beta_A^2)(1-\beta_B^2)} 
}.
\label{eq:f1}
\end{eqnarray}
Because this $\cos \theta_1$ distribution is P-even (see
Table~\ref{tab:cpt}), a non-trivial distribution requires parity
violation in both the production process ($a_W$) and in the decay of the
chargino ($a_C$).


In hadron collisions, event rates are obtained after a convolution with
the parton distribution functions. The actual distribution is $\propto 1
+ a_W \langle f_1 \rangle \cos \theta_1$ where $\langle f_1 \rangle$ is
an averaged value of $f_1$. The value of $f_1$ vanishes in the threshold
production limit ($\beta_A \to 0$, $\beta_B \to 0$), and it approaches
unity for a boosted event ($\beta_A \to 1$, $\beta_B \to 1$). (A larger
asymmetry can be observed if we select events with large $\hat s$,
although the number of events decreases exponentially if the lower cut
on $\hat s$ is increased.)
Observing this asymmetry will provide evidence of both the chargino spin
and of parity violation in the weak interaction of the
charginos and neutralinos.

\subsection*{\boldmath $\cos \theta_2$ distribution}

A similar distribution is obtained when we integrate $\theta$,
($\theta_1$, $\phi_1$), $\phi_2$, and $z_l$:
\begin{eqnarray}
  d\sigma &=& 
\sigma (q \bar q \to \chi^\pm \chi^0)
B(\chi^\pm \to \tilde{\tau}^\pm \nu)
B(\chi^0 \to \tilde{\tau}^\mp \tau^\pm)
B(\tau^\pm \to l^\pm \nu \bar \nu) {d \cos \theta_2 \over 2}
\nonumber \\
&& \times 
\left[ 1 + a_W a_N f_1(\beta_B, \beta_A) \cos \theta_2 \right].
\label{eq:cos2}
\end{eqnarray}
The angular dependence is due to both the neutralino spin and the parity
violation occurring both in the production and the decay of the
neutralino.

\subsection*{\boldmath $\cos \theta_1 \cos \theta_2$ distribution}

A non-trivial correlation is present between the angles on both sides of
decays. The $\theta_1$ and $\theta_2$ dependence of the cross section is
\begin{eqnarray}
  d\sigma &=& 
\sigma (q \bar q \to \chi^\pm \chi^0)
B(\chi^\pm \to \tilde{\tau}^\pm \nu)
B(\chi^0 \to \tilde{\tau}^\mp \tau^\pm)
B(\tau^\pm \to l^\pm \nu \bar \nu) 
{d \cos \theta_1 \over 2}
{d \cos \theta_2 \over 2}
\nonumber \\
&& \times 
[ 1 
+ a_W a_C f_1(\beta_A, \beta_B) \cos \theta_1 
+ a_W a_N f_1(\beta_B, \beta_A) \cos \theta_2 
\nonumber \\
&& \ \ \ \ \ 
+ a_C a_N f_2(\beta_A, \beta_B) \cos \theta_1 \cos \theta_2 
],
\label{eq:coscos2}
\end{eqnarray}
where
\begin{eqnarray}
 f_2 (\beta_A, \beta_B)
= \frac{
3 \beta_A \beta_B + 1 + \xi_W \sqrt{(1- \beta_A^2)(1-\beta_B^2)}
}{
3 + \beta_A \beta_B + 3\xi_W \sqrt{(1- \beta_A^2)(1-\beta_B^2)}
}.
\end{eqnarray}
Integrating over $\theta_1$ and $\theta_2$ keeping the product $\cos
\theta_1 \cos \theta_2 (\equiv y)$ fixed, we obtain
\begin{eqnarray}
  d\sigma &=& 
\sigma (q \bar q \to \chi^\pm \chi^0)
B(\chi^\pm \to \tilde{\tau}^\pm \nu)
B(\chi^0 \to \tilde{\tau}^\mp \tau^\pm)
B(\tau^\pm \to l^\pm \nu \bar \nu) {d y \over 2}
\nonumber \\
&& \times 
\left[ 1 + a_C a_N f_2(\beta_A, \beta_B) y \right]
\log | y |.
\label{eq:coscos}
\end{eqnarray}
The non-trivial part (the second term) is due to the spin correlations
between the chargino and the neutralino. Parity violation ($a_N \neq 0$)
biases the distribution towards a positive or negative value of $y$.
This distribution is independent of the asymmetry parameter $a_W$ in the
production process. Note also that the function $f_2$ does not vanish in
the limit of the threshold production ($f_2 \to 1/3$), although it is
maximized in the boost limit ($f_2 \to 1$). Confirming this correlation
will be an interesting test of the model.

\subsection*{\boldmath $\phi_1$ distribution}

A non-trivial distribution of the azimuthal angle $\phi_1$ takes
place due to parity violation in the chargino decay:
\begin{eqnarray}
  d\sigma &=& 
\sigma (q \bar q \to \chi^\pm \chi^0)
B(\chi^\pm \to \tilde{\tau}^\pm \nu)
B(\chi^0 \to \tilde{\tau}^\mp \tau^\pm)
B(\tau^\pm \to l^\pm \nu \bar \nu) {d \phi_1 \over 2 \pi}
\nonumber \\
&& \times 
\left[ 
1 
\pm { \pi^2 \over 16 } a_C g_1(\beta_A, \beta_B) \cos \phi_1 
\pm { \pi^2 \over 16 } a_C \eta_W g_2(\beta_A, \beta_B) \sin \phi_1 
\right],
\label{eq:phi1}
\end{eqnarray}
where
\begin{eqnarray}
 g_1 (\beta_A, \beta_B)
= {
\sqrt{1 - \beta_A^2} + \xi_W \sqrt{1- \beta_B^2}
\over
1 + \beta_A \beta_B / 3 + \xi_W \sqrt{(1- \beta_A^2)(1-\beta_B^2)}
},
\label{eq:g1}
\end{eqnarray}
\begin{eqnarray}
 g_2 (\beta_A, \beta_B)
= {
\beta_A \sqrt{1 - \beta_B^2} 
\over
1 + \beta_A \beta_B / 3 + \xi_W \sqrt{(1- \beta_A^2)(1-\beta_B^2)}
}.
\label{eq:g2}
\end{eqnarray}
The $\phi_1$ dependence appears even if $a_W = 0$. 
This is somewhat surprising once we realize the fact that $\cos \phi_1$
is P-even and the chargino decay violates parity ($a_C \neq 0$). In
order for the distribution to be formally P-invariant there should be
another interaction that violates parity. This is in fact supplied by
maximal parity violation in the weak interaction of the quarks in the
production process. This fact means that to observe the distribution one
needs to measure the direction of the quark (or the anti-quark). This
conclusion can be also seen in Fig.~\ref{fig:coordinate}, because
knowledge of the quark direction is necessary to define the angles
$\phi_1$ and $\phi_2$.
The different signs for the $\chi^+ \chi^0$ and $\chi^- \chi^0$
productions can be understood by the fact that $\cos \phi_1$ and $\sin
\phi_1$ are CPT-odd.
The $\sin \phi_1$ dependence (phase of the $\phi_1$ oscillation)
measures CP (or T) violation in the production process, $\eta_W$.

The function $g_1$ is maximized at the threshold limit and vanishes in
the boost limit, in contrast to the case of the polar-angle
dependencies. In the threshold limit, the coefficient of $\cos \phi_1$
is $\pi^2/16$. The CP asymmetry vanishes in both the threshold and the
boost limits.

The determination of the spin of the intermediate particles by looking
at the azimuthal-angle distributions (frequencies of the $\phi$
oscillations) has been discussed recently in
Ref.~\cite{Buckley:2007th}\footnote{As we have seen above in the case of
the $\chi^\pm \chi^0$ production, parity violation is needed at {\it
both} the production and the decay vertices in order to develop a
$\phi_1$ or $\phi_2$ azimuthal-angle dependence.
These conditions are in fact general requirements for $2 \to 2$ fermion
pair production with subsequent two-body decays of each fermion.
Therefore, the method of Ref.~\cite{Buckley:2007th} should work only in
a limited case.  For example, there is no azimuthal-angle dependence of
the differential cross section in processes where fermions are pair
produced through QED or QCD interactions such as $t \bar t$ pair
production or the production of a gluino pair at hadron colliders
(unless the beam is polarized). In such processes, the angular
correlations between two decays, such as the distribution of $\cos
\theta_1 \cos \theta_2$ or $\phi_1 \pm \phi_2$, will instead be useful
if there is parity violation at the decay vertices.}.

\subsection*{\boldmath $\phi_2$ distribution}

A similar distribution is obtained when $a_N$ is non-vanishing:
\begin{eqnarray}
  d\sigma &=& 
\sigma (q \bar q \to \chi^\pm \chi^0)
B(\chi^\pm \to \tilde{\tau}^\pm \nu)
B(\chi^0 \to \tilde{\tau}^\mp \tau^\pm)
B(\tau^\pm \to l^\pm \nu \bar \nu) {d \phi_2 \over 2 \pi}
\nonumber \\
&& \times 
\left[ 
1 
\mp { \pi^2 \over 16 } a_N g_1(\beta_B, \beta_A) \cos \phi_2 
\mp { \pi^2 \over 16 } a_N \eta_W g_2(\beta_B, \beta_A) \sin \phi_2 
\right].
\label{eq:phi2}
\end{eqnarray}

\subsection*{Other distributions}

Although we will not study them in this paper, there are various kinds
of other non-trivial distributions. For example, the distribution of the
difference of the angles $\phi_1 - \phi_2$ also depends on the
CP-violation parameter $\eta_W$. In the reconstruction of this angle at
hadron colliders we do not need to know the direction of the $q$ or
$\bar q$ in the initial state in contrast to the case of the angles
$\phi_1$ and $\phi_2$. This is an advantage especially at $pp$
colliders. Analytic formulae of such distributions can easily be
obtained from the full cross-section formula.

\section{LHC studies of \boldmath $\chi^+ \chi^-$ and $\chi^\pm \chi^0$ productions}

In this section, we demonstrate a possible strategy for the study of the
production processes of charginos and neutralinos by performing a Monte
Carlo simulation. We use the following simplified model for generating
events:
\begin{eqnarray}
 \mu = 300~{\rm GeV}, \ \ \ 
M_1 = M_2 = m_{\tilde{u}_L}
= m_{\tilde{d}_L}= 5000~{\rm GeV},
\end{eqnarray}
\begin{eqnarray}
 \tan \beta = 10,\ \ \ 
 m_{\tilde{\tau}_L} = 5000~{\rm GeV}, \ \ \ 
 m_{\tilde{\tau}_R} = 100~{\rm GeV}.
\end{eqnarray}
With this choice of parameters, all the SUSY particles decouple from low
energy except for the Higgsinos and the right-handed $\tilde \tau$. The
chargino and two light neutralinos are purely Higgsino-like and the
masses are calculated to be:
\begin{eqnarray}
 m_{\tilde{\tau}_1} = 109~{\rm GeV}, \ \ \ 
 m_{\chi^+_1} = 300~{\rm GeV}, \ \ \ 
 m_{\chi^0_1} = 299~{\rm GeV}, \ \ \ 
 m_{\chi^0_2} = 301~{\rm GeV}.
\end{eqnarray}
Although there are two mass eigenstates for the neutralinos due to a
small mixing with gauginos, they almost behave like a single Dirac
fermion. We do not distinguish $\chi_1^0$ and $\chi_2^0$ in the
following analysis.
The lifetime of $\tilde \tau_1$ is assumed to be much longer than the
typical collider time scale ($1/\Gamma_{\tilde \tau} \gg$~ns).
The branching fractions of the chargino and the neutralino decays are of
course,
\begin{eqnarray}
 B( \chi^\pm \to \tilde \tau^\pm \nu ) \simeq 1.0,\ \ \ 
 B( \chi^0 \to \tilde \tau^\pm \tau^\mp ) \simeq 0.5.
\end{eqnarray}
The asymmetry parameters in this model are calculated to be
\begin{eqnarray}
 a_N = 1.00,\ \ \ 
 a_W = 0.00,\ \ \
 \xi_W = 1.00,\ \ \ 
 \eta_W = 0.00.
\end{eqnarray}
Note that this parameter choice is simply for a demonstration and not
particularly motivated by any fundamental model which realizes a light
$\tilde \tau$. We use this model as the first trial of the study of
production events of charginos and neutralinos in the long-lived $\tilde
\tau$ scenario. In more realistic situations, other mass eigenstates
will be produced which contaminates the analysis of the leading
production process. Since the importance of such effects depends on the
detailed structure of the models, we use the above clean model as a toy
example.
The values of asymmetry parameters, $a_W = 0.00$ and $\eta_W = 0.00$,
are not very interesting ones, but in fact, as we see later we need to
first study these trivial cases in order to confirm whether there is a
fake distribution caused by false solutions, which appear in the
reconstruction of kinematic variables. In order to measure the
asymmetry, we need to understand whether a non-trivial distribution is
fake or physical.

We have generated 26,000 events of the electroweak production processes
of SUSY particles (including the $\tilde \tau$ pair-production process)
in the $pp$ collision at $\sqrt s = 14$~TeV by using the Herwig~6.5
event generator~\cite{Corcella:2002jc}. The spin correlations have been
implemented for the $\chi^+ \chi^0$ production and their
decays~\cite{Richardson:2001df, Moretti:2002eu}.  This number of events
corresponds to an integrated luminosity of 100~fb$^{-1}$ at the LHC. (We
will use 300~fb$^{-1}$ of data for some of the analysis of angular
distributions.) We have used the CTEQ5L library~\cite{Lai:1999wy} for
the parton distribution function.
For the $\tau$ decay, we have used TAUOLA~2.7
package~\cite{Jadach:1993hs} so that the spin information is maintained.
A detector simulator AcerDET~1.0~\cite{RichterWas:2002ch} has been used
for the event analysis.

In the following analysis, we assume that the mass of $\tilde \tau$ is
known by the method of Ref.~\cite{Ambrosanio:2000ik}, and we ignore the
resolution of the $\tilde \tau$-momentum measurements which is of order
a few percent in the ATLAS experiment~\cite{stau}. One should note that
the accuracies of the measurement quoted below are somewhat optimistic
for this reason. We also assume perfect efficiencies of the $\tilde
\tau$ identification and of the $\tilde \tau$-charge measurement for
$\tilde \tau$ tracks with $p_T > 10$~GeV and $|\eta| < 2.5$. By
requiring two $\tilde \tau$'s, there is no Standard Model background
with this assumption although in actual experiments one needs to take
into account mis-identifications of muons as $\tilde \tau$.

We first discuss possible methods to measure the masses of the chargino
and the neutralino through the exclusive production
processes. Measurements of the asymmetries from looking at the angular and
energy distributions studied in Section~\ref{sec:dist} will then be
demonstrated.

\subsection{Chargino mass determination by chargino-pair production}

\begin{figure}[t]
 \begin{center}
 \includegraphics[width=7.5cm]{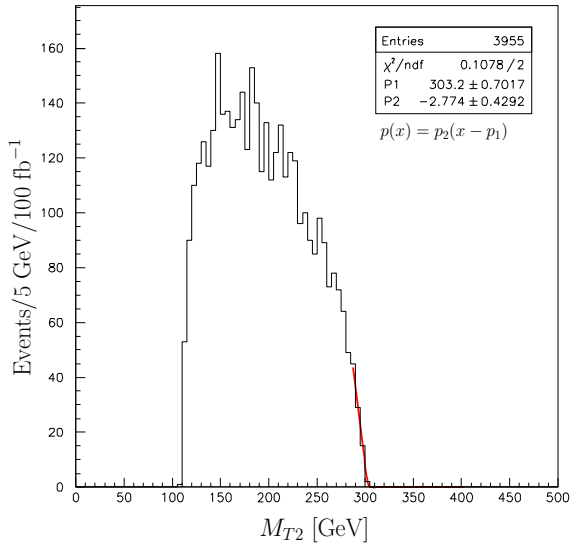}
\end{center}
\caption{The $M_{T2}$ distribution of the chargino-pair production. The
 chargino mass can be extracted by looking at the endpoint.}
\label{fig:mT2}
\end{figure}

We present a method to measure the chargino mass exclusively from the
chargino pair-production process. The final state of the process is two
opposite-sign $\tilde \tau$'s and missing momentum from the two
neutrinos.

Although we cannot reconstruct the chargino momentum on an
event-by-event basis, the endpoint analysis developed in
Ref.~\cite{Lester:1999tx} can be used to extract the chargino mass. The
method is to form a quantity $M_{T2}$ defined by
\begin{eqnarray}
 M_{T2}^2 = \min_{{\bf p}_{T\nu_1}+{\bf p}_{T\nu_2} = {\bf
  p}_T^{\rm miss}}
\left[
\max \{
m_T^2({\bf p}_{T \tilde \tau^-}, {\bf p}_{T \nu_1}),
m_T^2({\bf p}_{T \tilde \tau^+}, {\bf p}_{T \nu_2})
\}
\right],
\end{eqnarray}
where ${\bf p}_{T \tilde \tau^-}$ and ${\bf p}_{T \tilde \tau^+}$ are
the transverse momentum of the two $\tilde \tau$'s and ${\bf p}_T^{\rm
miss}$ is the missing transverse momentum. The transverse mass $m_T$ is
defined by
\begin{eqnarray}
 m_T^2 &=& (E_{T \tilde{\tau}} + E_{T \nu})^2 
- | {\bf p}_{T \tilde{\tau}} +  {\bf p}_{T \nu} |^2 
\nonumber \\
&=& m_{\tilde{\tau}}^2 + 2
( E_{T \tilde{\tau}} E_{T \nu} 
- {\bf p}_{T \tilde{\tau}} \cdot {\bf p}_{T \nu} ),
\label{eq:trans}
\end{eqnarray}
where
\begin{eqnarray}
 E_T^2 = m^2 + |{\bf p}_T|^2.
\end{eqnarray}
The quantity $M_{T2}$ is designed to have the endpoint at the mass of
the intermediate particle.

In order to select the $\chi^+ \chi^-$ events, we have imposed the
following jet and lepton vetoes:
\begin{eqnarray}
N_{\tilde \tau^+} = N_{\tilde \tau^-} = 1,\ \ \ 
  N_j (p_{T} > 30~{\rm GeV})= 0,\ \ \
 N_l (p_{T} > 6~{\rm GeV}) = 0.
\end{eqnarray}
We do not need to impose a tight cut on the missing momentum since
$\tilde \tau^+ \tilde \tau^-$ events do not contribute near the endpoint
of the distribution.
The $M_{T2}$ distribution is shown in Fig.~\ref{fig:mT2}. There is a
clear endpoint around the input chargino mass, 300~GeV. By fitting with
a linear function, we obtain the endpoint: $303.2 \pm 0.7~{\rm GeV}$,
which is slightly larger than the input value due to the resolution of
the missing transverse momentum\footnote{One should use a
Gaussian-smeared line to take into account the effect of finite
resolutions.}.

\subsection{Neutralino and chargino mass determination by chargino-neutralino production}

We show in this subsection that quite accurate measurements of the
neutralino and chargino masses are possible by analyzing exclusive
processes. Combining various methods described below, we will be able to
measure the masses at the level of a few GeV.

\subsubsection{Endpoint analysis for the neutralino mass}

\begin{figure}[t]
\begin{center}
 \includegraphics[width=7.5cm]{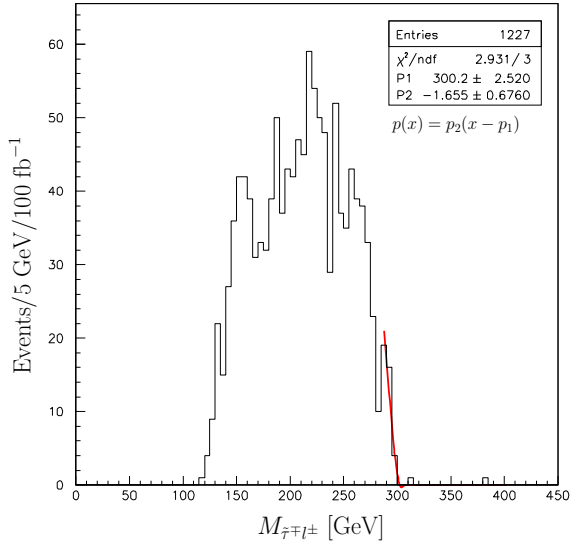}
\end{center}
\caption{Invariant mass distribution of $\tilde \tau^\mp l^\pm$
 pairs. The endpoint shows the neutralino mass.}  \label{fig:minv}
\end{figure}

The neutralino mass can be measured by looking for an endpoint of the
invariant mass distribution of the $\tilde \tau^\mp l^\pm$ pair from the
neutralino decay followed by the leptonic tau decay. The $\chi^\pm
\chi^0$-production process can be selected by requiring two $\tilde
\tau$'s and an isolated lepton:
\begin{eqnarray}
 N_{\tilde \tau^+} = N_{\tilde \tau^-} = 1,\ \ \ 
 N_j (p_{T} > 30~{\rm GeV})= 0,\ \ \
 N_l (p_T > 10~{\rm GeV})= 1.
\label{eq:cut}
\end{eqnarray}
As we discussed before, we require that two $\tilde \tau$'s have
opposite signs so that there is no ambiguity in selecting $\tilde \tau$
from the neutralino decay.

The invariant mass distribution of the $\tilde \tau^\mp l^\pm$ pair is
shown in Fig.~\ref{fig:minv}. An accurate measurement of the neutralino
mass is possible by this method ($300 \pm 3$~GeV).

\subsubsection{Solvability analysis for the neutralino mass}

\begin{figure}[t]
\begin{center}
 \includegraphics[width=7.6cm]{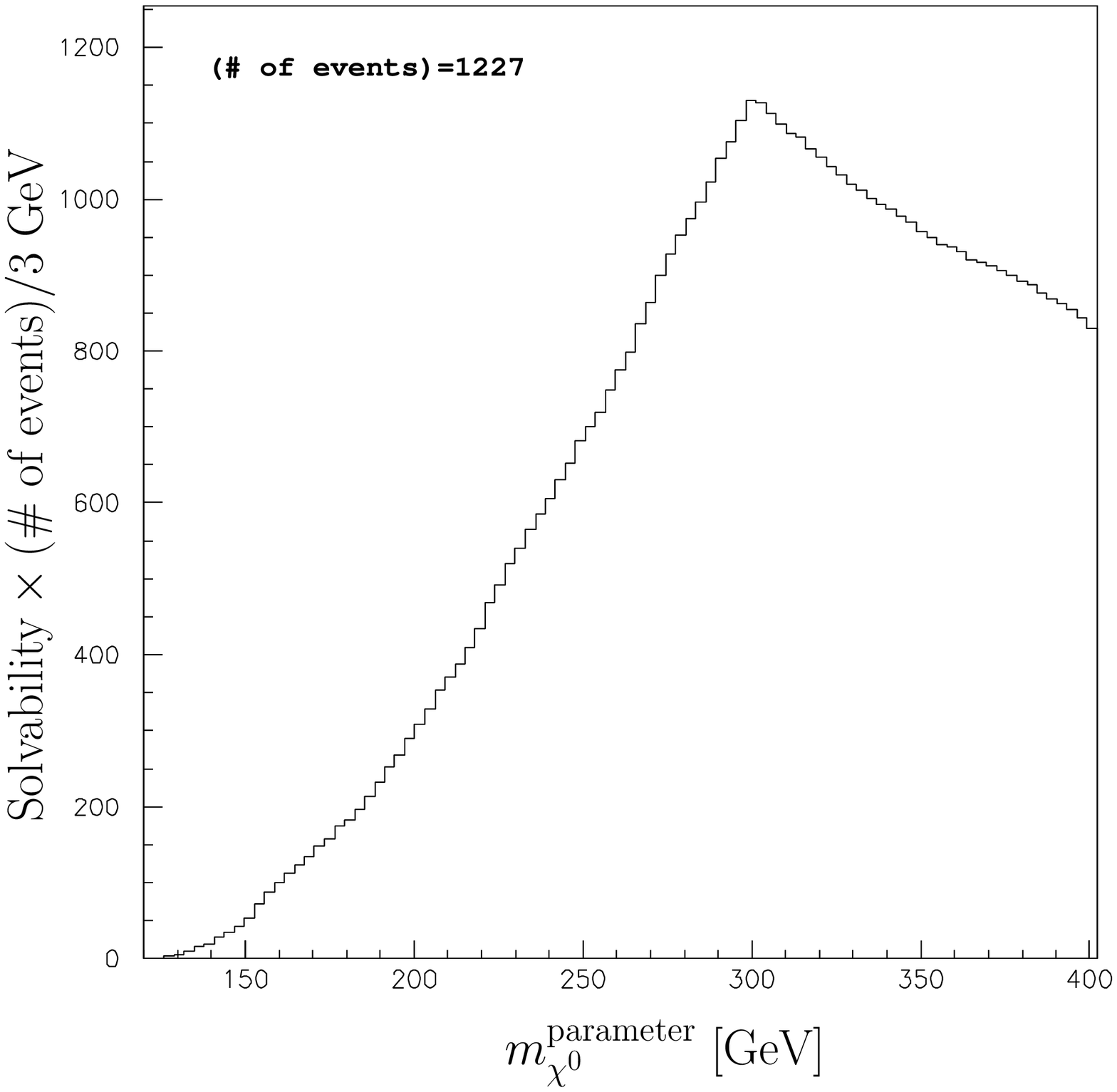}
 \includegraphics[width=7.5cm]{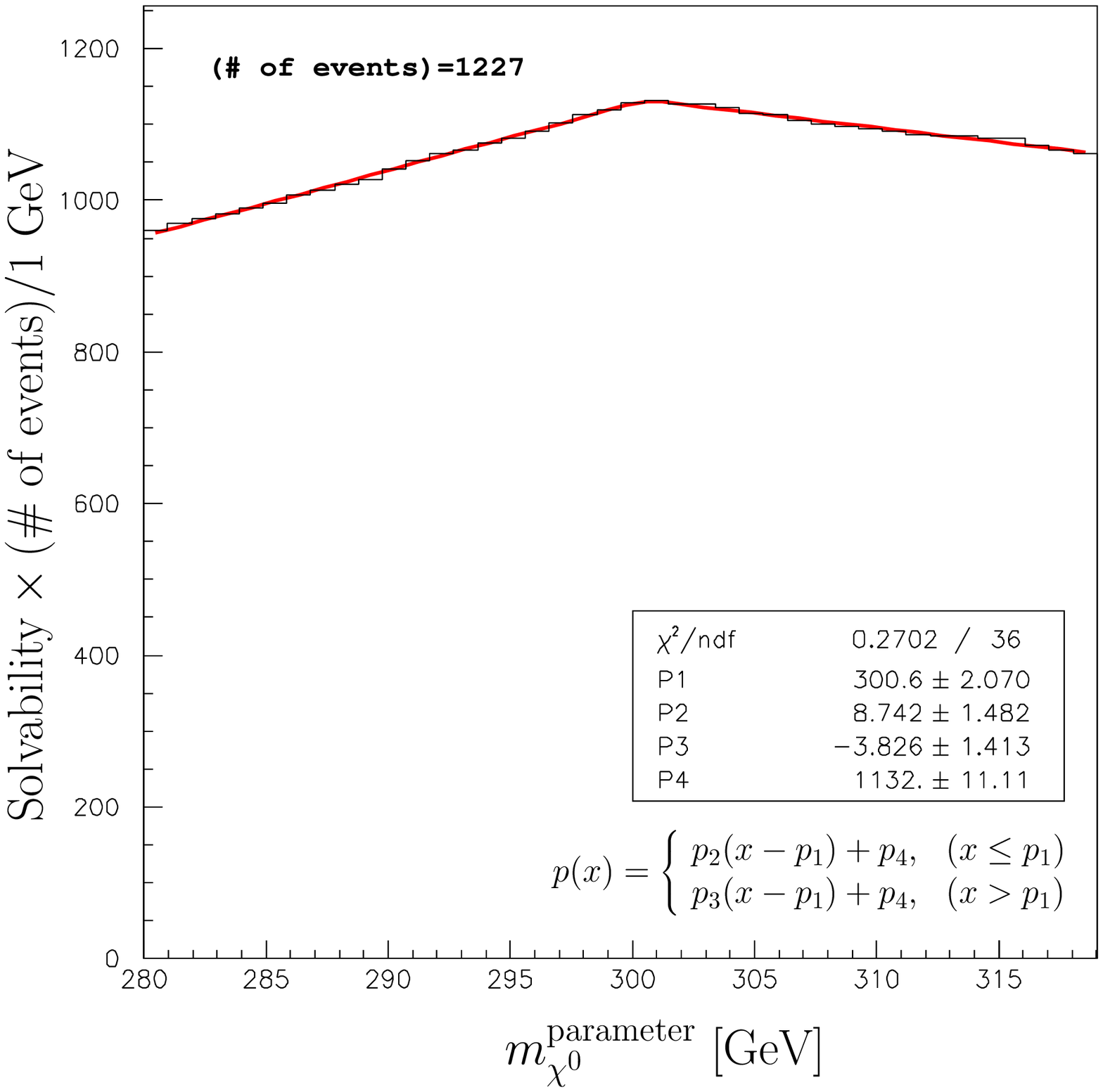}
\end{center}
\caption{The solvability analysis for the neutralino mass. The right
 figure shows the solvability near the peak.} 
\label{fig:solv-neut}
\end{figure}

By using the information of the chargino mass measured by the $\chi^+
\chi^-$ pair production process, we can obtain the neutralino mass by a
similar method proposed in Refs.~\cite{Kawagoe:2004rz,
Cheng:2007xv}. (See also \cite{CMS} for a similar analysis for the
measurement of the top-quark mass in the di-lepton events from the $t
\bar t$ productions at the LHC.)
Since the final state is relatively simple, we can solve the kinematics
on an event-by-event basis by postulating a neutralino mass. By
maximizing the solvability (number of events which can give physical
solutions normalized by the total number of events analyzed), we can
obtain the correct neutralino mass.

The equations to be satisfied are
\begin{eqnarray}
 (P_{\tilde{\tau}^\pm} + P_{\nu})^2 = m_{\chi^+}^2,
\label{eq:pz}
\end{eqnarray}
\begin{eqnarray}
 (P_{\tilde{\tau}^\mp} + {P_{l^\pm} \over z_l}  )^2 = m_{\chi^0}^2,
\label{eq:zeq}
\end{eqnarray}
\begin{eqnarray}
 P_\nu^x + {1 - z_l \over z_l } P_{l^\pm}^x = P_{\rm miss}^x,
\label{eq:px}
\end{eqnarray}
\begin{eqnarray}
 P_\nu^y + {1 - z_l \over z_l } P_{l^\pm}^y = P_{\rm miss}^y,
\label{eq:py}
\end{eqnarray}
whereas there are four unknowns in these equations:
\begin{eqnarray}
 P_\nu^x,\ \ \  P_\nu^y,\ \ \  P_\nu^z,\ \ \ z_l.
\end{eqnarray}
Equation~(\ref{eq:zeq}) is a linear equation for $z_l$ in the
approximation of $m_l = 0$, and by using the solution,
Eqs.~(\ref{eq:px}) and (\ref{eq:py}) become also linear equations for
$P_\nu^x$ and $P_\nu^y$, respectively. Eq.~(\ref{eq:pz}), on the other
hand, is a quadratic equation for $P_\nu^z$, and therefore there can be
either zero or two real-number solutions. (In general, an equation of
this type may have a unique physical solution by the constraint $E_\nu
>0$. However, one can show that Eq.~(\ref{eq:pz}) always have zero or
two solutions by using the fact that $E_{\tilde \tau^\pm} > |P_{\tilde
\tau^\pm}^z|$.)
The number of events should be maximized at the
correct neutralino mass when we impose conditions: $0 \leq z_l \leq 1$
and existence of real-number solutions of Eq.~(\ref{eq:pz}).

The number of events with a physical solution is shown in
Fig.~\ref{fig:solv-neut} for various input neutralino masses. It indeed
shows a sharp peak at the correct neutralino mass, 300~GeV. We have used
the correct value of the chargino mass in the analysis. In the actual
situation, the experimental error in the chargino mass will propagate
into the error in the peak location.
Fitting with two linear functions near the peak, we find that the
neutralino mass can be measured quite accurately ($301 \pm 2$~GeV) if
the chargino mass is known.

Note again that this analysis is not completely realistic. We have
ignored the momentum resolutions of $\tilde \tau$ tracks and assumed the
perfect identification efficiency. We leave more realistic studies to
future work.

\subsubsection{Transverse mass analysis for the chargino mass}

\begin{figure}[t]
\begin{center}
 \includegraphics[width=7.5cm]{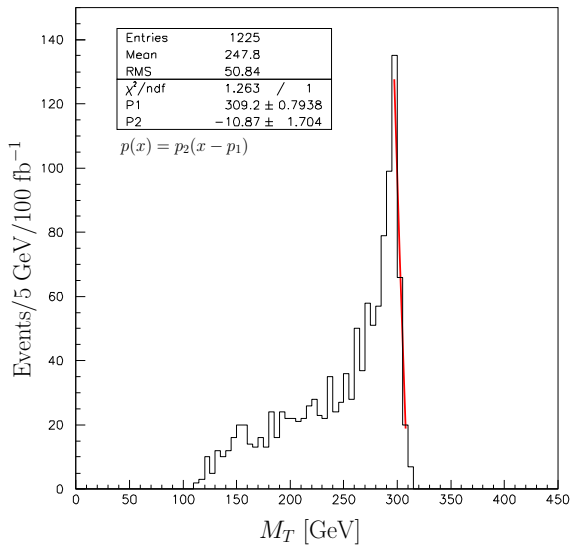}
\end{center}
\caption{The transverse mass distribution for the chargino mass measurement.} 
\label{fig:mT}
\end{figure}

Once we know the neutralino mass by, for example, the method of the
endpoint of the $M_{\tilde \tau^\mp l^\pm}$ distribution, the transverse
momentum of the neutrinos from the chargino decay can be reconstructed
without the two-fold ambiguity from Eqs.~(\ref{eq:zeq}--\ref{eq:py}). We
can then form a transverse mass in Eq.~(\ref{eq:trans}) and the chargino
mass can be obtained by looking for an endpoint of the distribution.

We show in Fig.~\ref{fig:mT} the distribution of the transverse mass,
$M_T$. We can see a sharp peak near the correct chargino mass. The
endpoint is again smeared by the resolution of the missing transverse
momentum. An appropriate fitting is necessary for the extraction of the
chargino mass. For a simple fitting by a linear function, we obtain a
significantly larger value ($309.2\pm0.8$~GeV) due to the finite
resolution.

\subsubsection{Solvability analysis for the chargino mass}

\begin{figure}[t]
\begin{center}
 \includegraphics[width=7.5cm]{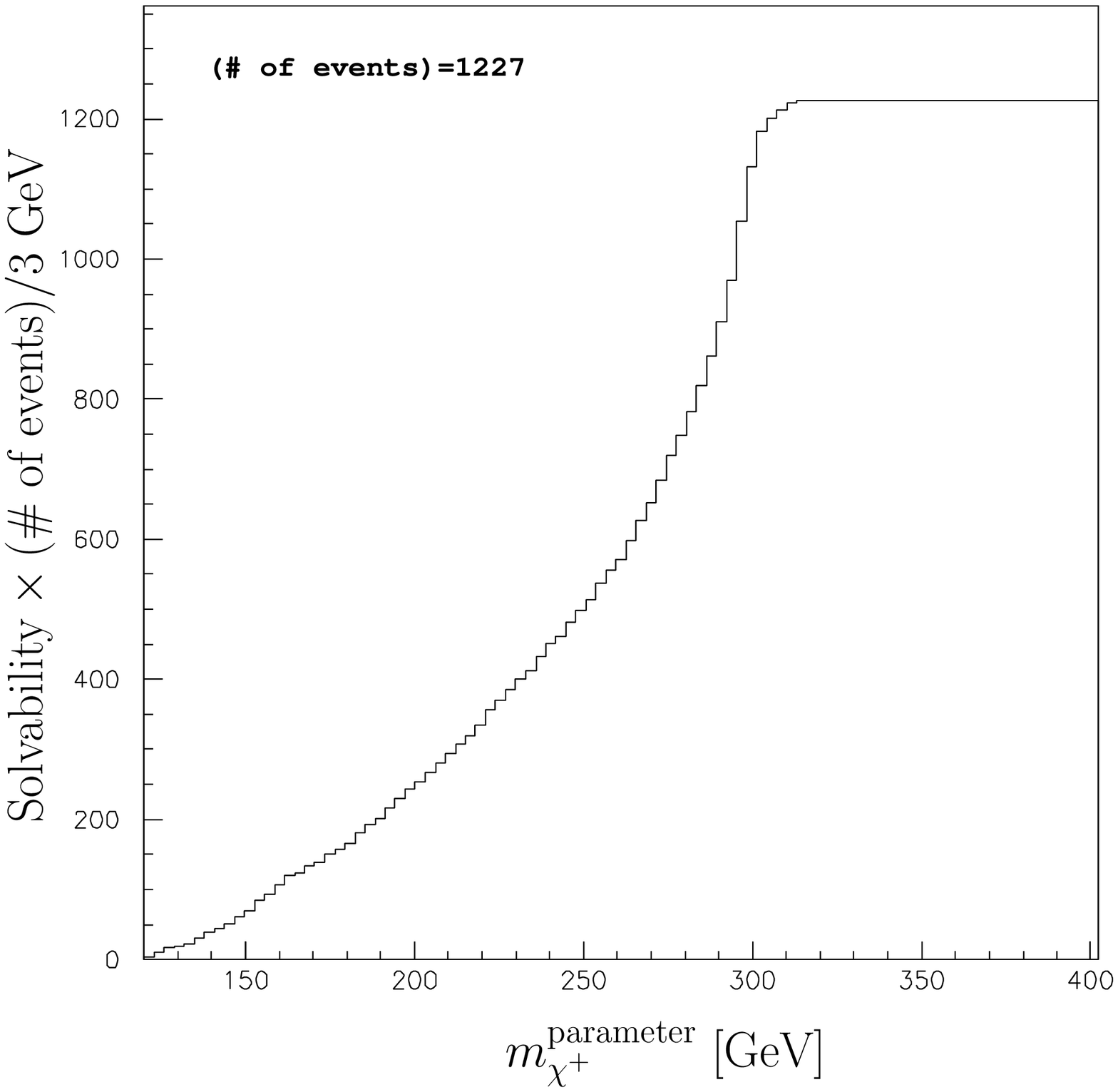}
 \includegraphics[width=7.5cm]{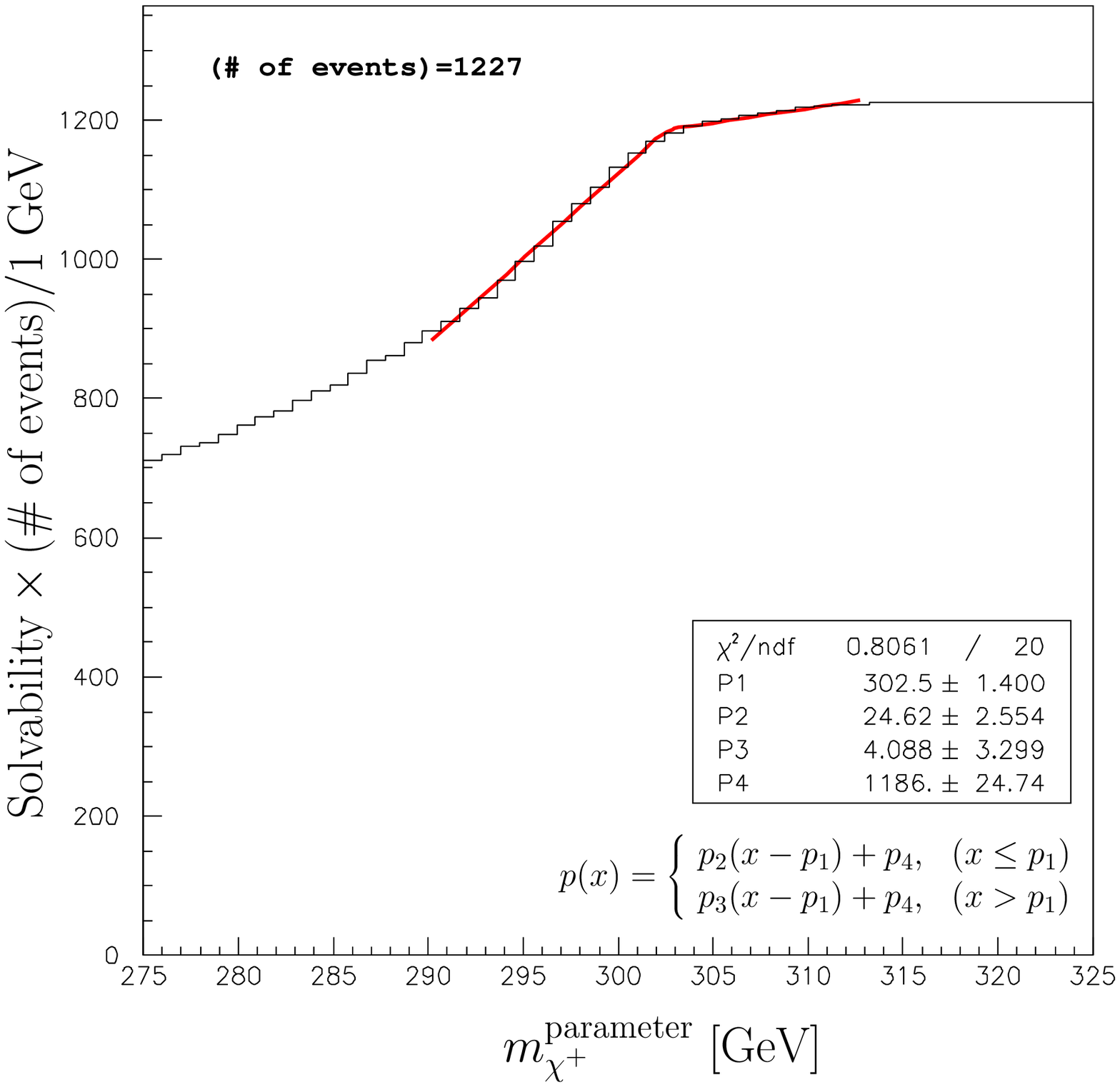}
\end{center}
\caption{The solvability analysis for the chargino mass
 measurement. The right figure is the same analysis with a better
 resolution near the threshold.}  \label{fig:solv-char}
\end{figure}

The solvability analysis can also be done for the chargino mass once we
know the neutralino mass. The solvability is plotted in
Fig.~\ref{fig:solv-char} where we see that the solvability saturates
near the chargino mass.

By looking for a point where the solvability saturates, we can obtain
the chargino mass ($303\pm1$~GeV).

\subsection{Energy and angular distributions}

Now we examine whether the energy and angular distributions obtained in
Section~\ref{sec:dist} are visible in actual experiments. An especial
concern is that there is always a false solution in the
Eqs.~(\ref{eq:pz}--\ref{eq:py}), which may destroy the theoretical
distributions. One purpose of this analysis is to understand the effect
of the false solution.

In the analysis, we have used the events passed through the selection
cut in Eq.~(\ref{eq:cut}).
We assume in the following that the chargino and neutralino masses are
known and ignore errors in the mass measurements.

\subsubsection{\boldmath $z_l$ distribution}

\begin{figure}[t]
\begin{center}
 \includegraphics[width=7.5cm]{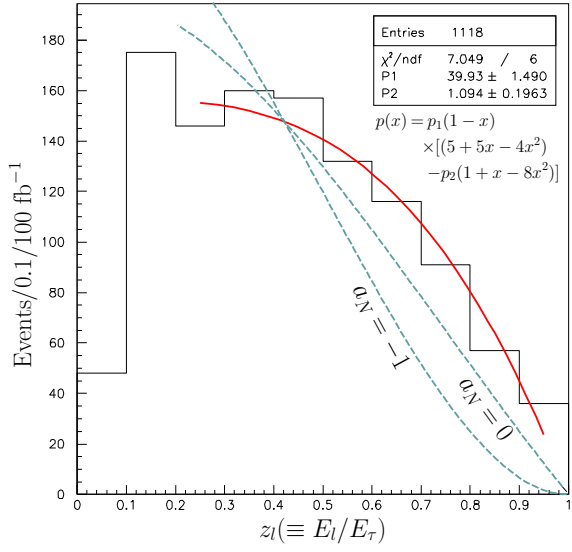}
\end{center}
\caption{The reconstructed distribution of the lepton-energy fraction
 ($z_l$) . The solid curve is the best fit with a theoretical function
 in Eq.~(\ref{eq:z_l}). The curves with $a_N = 0$, $-1$ are also shown
 (dashed curves).}  \label{fig:z-lepton}
\end{figure}

This distribution measures the polarization of the $\tau$ lepton from
the neutralino decay. We do not need to distinguish events with
different lepton charges since the theoretical distributions are the
same (Eq.~(\ref{eq:z_l})).

The measurement of the energy fraction of the lepton, $z_l$, does not
suffer from the two-fold ambiguity since Eq.~(\ref{eq:zeq}) is a linear
equation in $z_l$ in the approximation of $m_l = 0$, and we do not need
to know $P_\nu^z$.

The distribution is shown in Fig.~\ref{fig:z-lepton}. We fit the
distributions with the theoretical curve in Eq.~(\ref{eq:z_l}) by making
$a_N$ a parameter. We obtain $a_N = 1.1 \pm 0.2$ (solid curve) for
$a_N^{\rm th} = 1.0$ in this model. Curves with $a_N = 0$ and $a_N = -1$
are shown in the figure (dashed lines, we used the same normalization
with the solid curve). We can see that the best-fit curve can be
discriminated from those models.
The region with small $z_l$ is affected by the $p_T$ cut on the
leptons. This region is omitted from the fitting.

\subsubsection{\boldmath $\cos \theta_1$ distribution}

\begin{figure}[t]
\begin{center}
 \includegraphics[width=7.5cm]{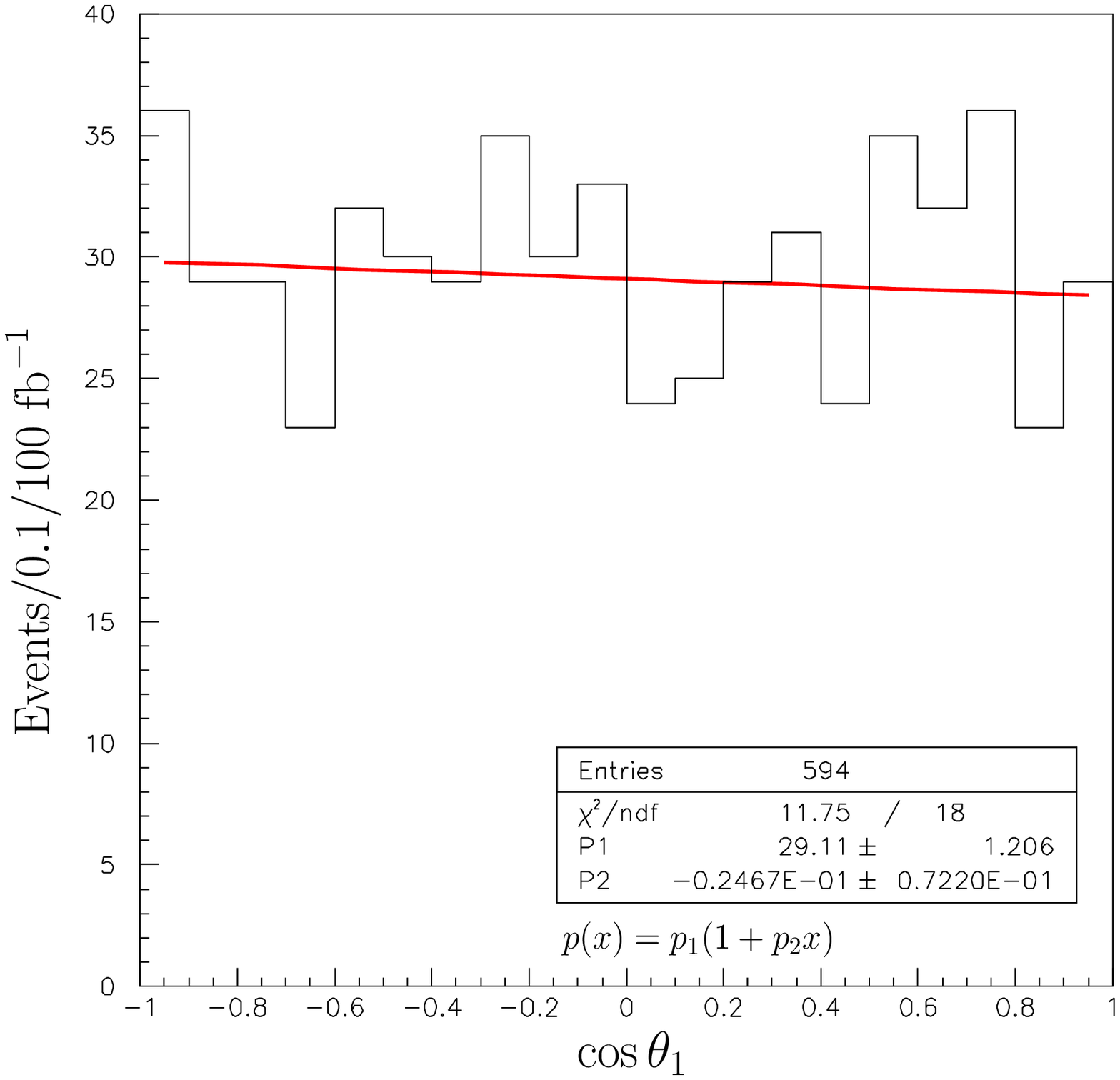}
 \includegraphics[width=7.5cm]{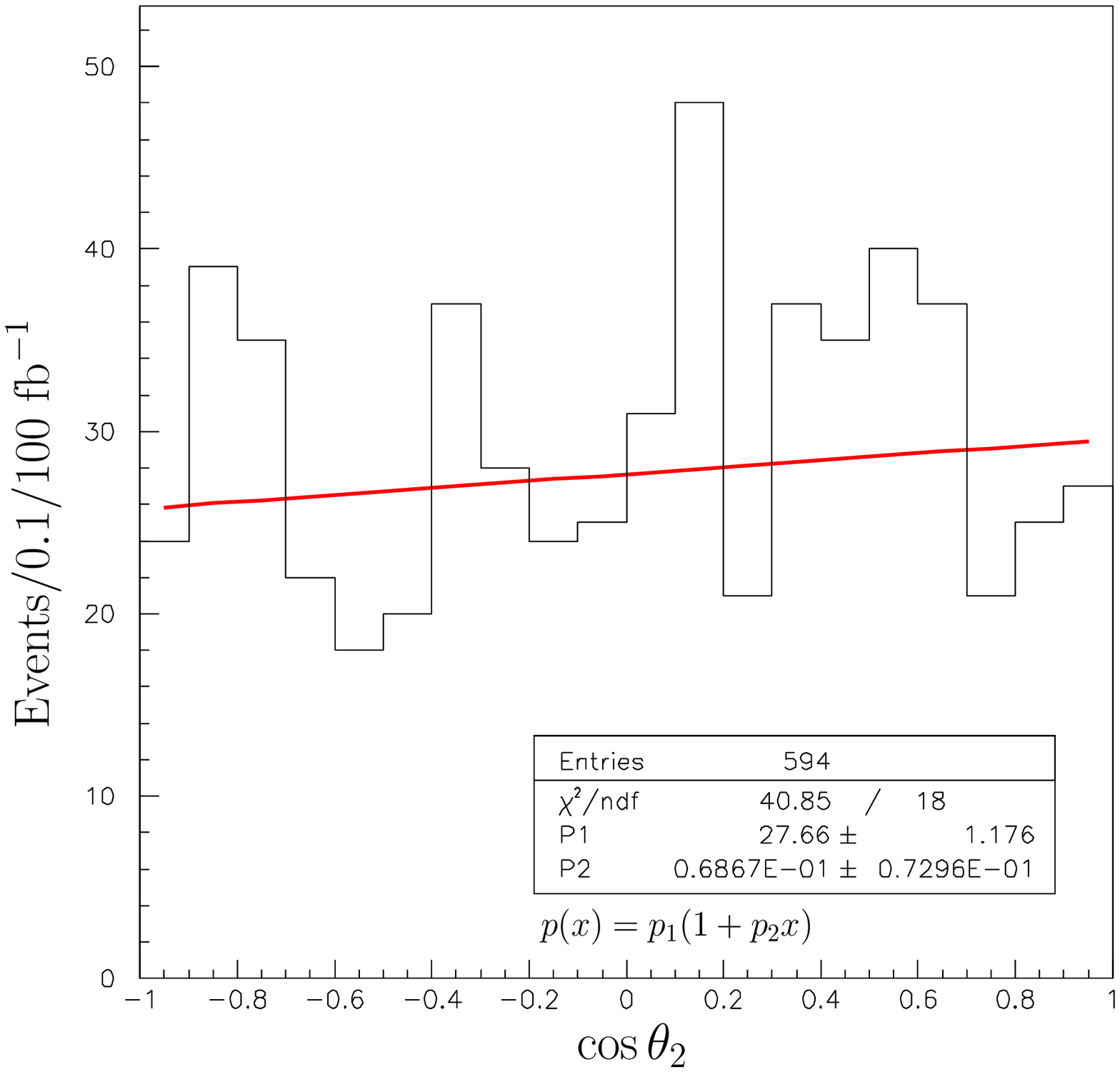}
\end{center}
\caption{The reconstructed $\cos \theta_1$ (left) and $\cos \theta_2$
(right) distributions. The false solutions are included. A selection cut
$\sqrt{\hat s} > 900$~GeV is imposed on both of the solutions in each
event.}  \label{fig:cos1}
\end{figure}

This distribution measures the parity asymmetry in the $\chi^\pm \chi^0$
production process, $a_W$, through Eq.~(\ref{eq:cos1}). The averaged
value of the function $f_1$ in Eq.~(\ref{eq:f1}) weighted by the cross
section depends on a selection cut on $\hat s$ ($\hat s \equiv
(P_{\chi^\pm} + P_{\chi^0})^2$). It is an increasing
function of $\hat s_{\rm min}$, but the number of events rapidly
decreases with $\hat s_{\rm min}$. In the model we simulated, the cross
section falls off as
\begin{eqnarray}
{ d \sigma (\hat s) \over d \sqrt{\hat s}} \propto 
\exp \left[ - 4.3 \left( {\sqrt{\hat s} \over {\rm TeV}} \right)
\right].
\label{eq:s-dep}
\end{eqnarray}
The averaged value defined by
\begin{eqnarray}
\langle f_1 (\sqrt{\hat{s}}_{\rm min}) \rangle
\equiv
 {
\displaystyle{\int_{\sqrt{\hat{s}}_{\rm min}}^{\infty}} 
{\displaystyle{ d \sigma ({\hat s}) \over d \sqrt{\hat s} }} 
f_1(\sqrt{\hat s}) d \sqrt{\hat s}
\over
\displaystyle{\int_{\sqrt{\hat{s}}_{\rm min}}^{\infty}} 
\displaystyle{d \sigma (\hat s) \over d \sqrt{\hat s}} 
d \sqrt{\hat s}
} 
\end{eqnarray}
is then estimated to be, for example,
\begin{eqnarray}
 \langle f_1 (600~{\rm GeV})  \rangle = 0.52
,\ \ \ 
 \langle f_1 (900~{\rm GeV}) \rangle  = 0.74,
\end{eqnarray}
for $m_{\chi^+} \simeq m_{\chi^0} \simeq 300$~GeV and $\xi_W = 1$.
Since the asymmetry will be diluted by false solutions as we see below,
it is necessary to impose a cut $\hat s_{\rm min}$ in order to expect a
large asymmetry.

The angle $\cos \theta_1$ which is defined in the rest frame of the
chargino is expressed in terms of the $\tilde \tau^\pm$ energy in the CM
frame:
\begin{eqnarray}
 \cos \theta_1 = 
{ 
E_{\tilde{\tau}^{\pm}}^{\rm CM} - \gamma_A E_{\tilde{\tau}^\pm}^{\rm {
(1)}}
\over 
\gamma_A \beta_A P_{\tilde{\tau}^\pm}^{\rm (1)}
},
\end{eqnarray}
where $\gamma_A$ and $\beta_A$ are defined in Eq.~(\ref{eq:gammaA}), and
$E_{\tilde \tau^\pm}^{(1)}$ and $P_{\tilde \tau^\pm}^{(1)}$ are the
energy and momentum in the rest frame of the chargino,
respectively. They are given by
\begin{eqnarray}
 E_{\tilde{\tau}^\pm}^{(1)} = 
{
m_{\chi^+}^2 + m_{\tilde \tau}^2
\over
2 m_{\chi^+}
}, \ \ \ 
P_{\tilde{\tau}^\pm}^{(1)} =
\sqrt{
\left( E_{\tilde{\tau}^\pm}^{(1)} \right)^2 - m_{\tilde \tau}^2
}.
\end{eqnarray}
When we calculate the $\tilde \tau$ energy in the CM frame from the
quantities measured in the laboratory frame by boosting to the
$z$-direction, one encounters the two-fold ambiguity for $P^z_\nu$ in
Eq.~(\ref{eq:pz}).

In order not to develop a fake distribution caused by the false
solution, we need to be careful when we impose a selection cut on $\hat
s$.
We have examined the following three methods of imposing a $\hat
s_{\rm min}$ cut.
One is to choose a solution which gives a smaller value of $\hat s$, and
impose a selection cut $\sqrt{\hat s}>900$~GeV on the chosen event. This
strategy effectively picks up the true solution (with the probability of
about 63\%) because of the distribution in Eq.~(\ref{eq:s-dep}).
However, this method causes a bias in the $\cos \theta_1$ distribution
towards larger $\cos \theta_1$.
For each event, it is likely that the solution with larger $\cos
\theta_1$ (which would mean that the neutrino is emitted to the opposite
direction to the chargino in the CM frame) gives a smaller value of
$\hat s$, and thus such a solution is more probable to be chosen. This
correlation causes bias.

The next strategy is to use all the solutions with $\sqrt{\hat s} >
900$~GeV. That is, if we have two solutions which satisfy the cut in an
event, we use both solutions. If there is only one solution with
$\sqrt{\hat s}>900$~GeV, we use that one. This strategy causes a fake
distribution towards smaller $\cos \theta_1$ this time. Since we impose
a lower cut on $\hat s$, this strategy tends to select a solution with
larger $\hat s$. By the same reason as above, this tends to pick up a
solution with smaller $\cos \theta_1$.

The above two lessons lead us to a good strategy to avoid the bias. It
is to use both solutions in each event and impose an $\hat s$ cut on
{\it both} of the solutions, i.e., we throw away an event if there is a
solution with $\sqrt{\hat s} < 900$~GeV even though it may be a false
solution. By doing that, the probability of selecting the true solution
is exactly 50\%, and there is no obvious reason to expect a fake
distribution.
We show in Fig.~\ref{fig:cos1} the reconstructed $\cos \theta_1$
distribution by using the strategy (left figure). A flat distribution is
obtained which is expected in this model because $a_W = 0.0$. For a more
general case, the slope of this distribution should give approximately
$a_W \langle f_1 \rangle / 2$ where the factor of two comes from the
effect of false solutions.
As the statistical error of $p_2$ in Fig.~\ref{fig:cos1} is $\pm 0.07$,
the establishment of $a_W \neq 0$ at the $3\sigma$ level would require $|a_W|
> 0.6$.

\subsubsection{\boldmath $\cos \theta_2$ distribution}

This measures the product of the parity asymmetries $a_N$ and $a_W$ in
Eq.~(\ref{eq:cos2}) through the spin correlation of the neutralino.
By the same strategy as in the $\cos \theta_1$ case, we obtain a flat
distribution for $\cos \theta_2$ as expected. It is shown in the right
panel of Fig.~\ref{fig:cos1}.

\subsubsection{\boldmath $\cos \theta_1 \cos \theta_2$ distribution}

\begin{figure}[t]
\begin{center}
 \includegraphics[width=7.5cm]{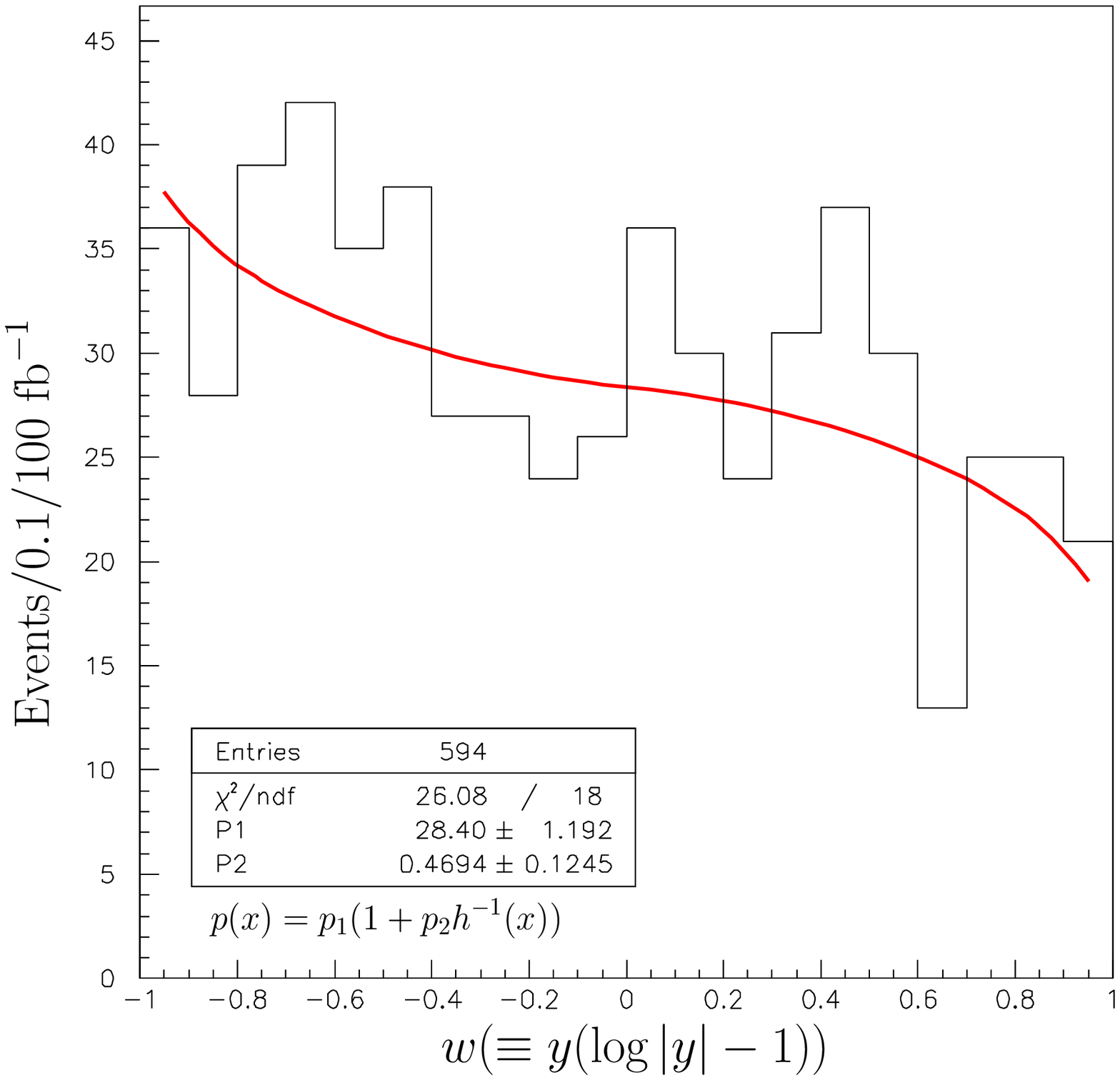}
 \includegraphics[width=7.5cm]{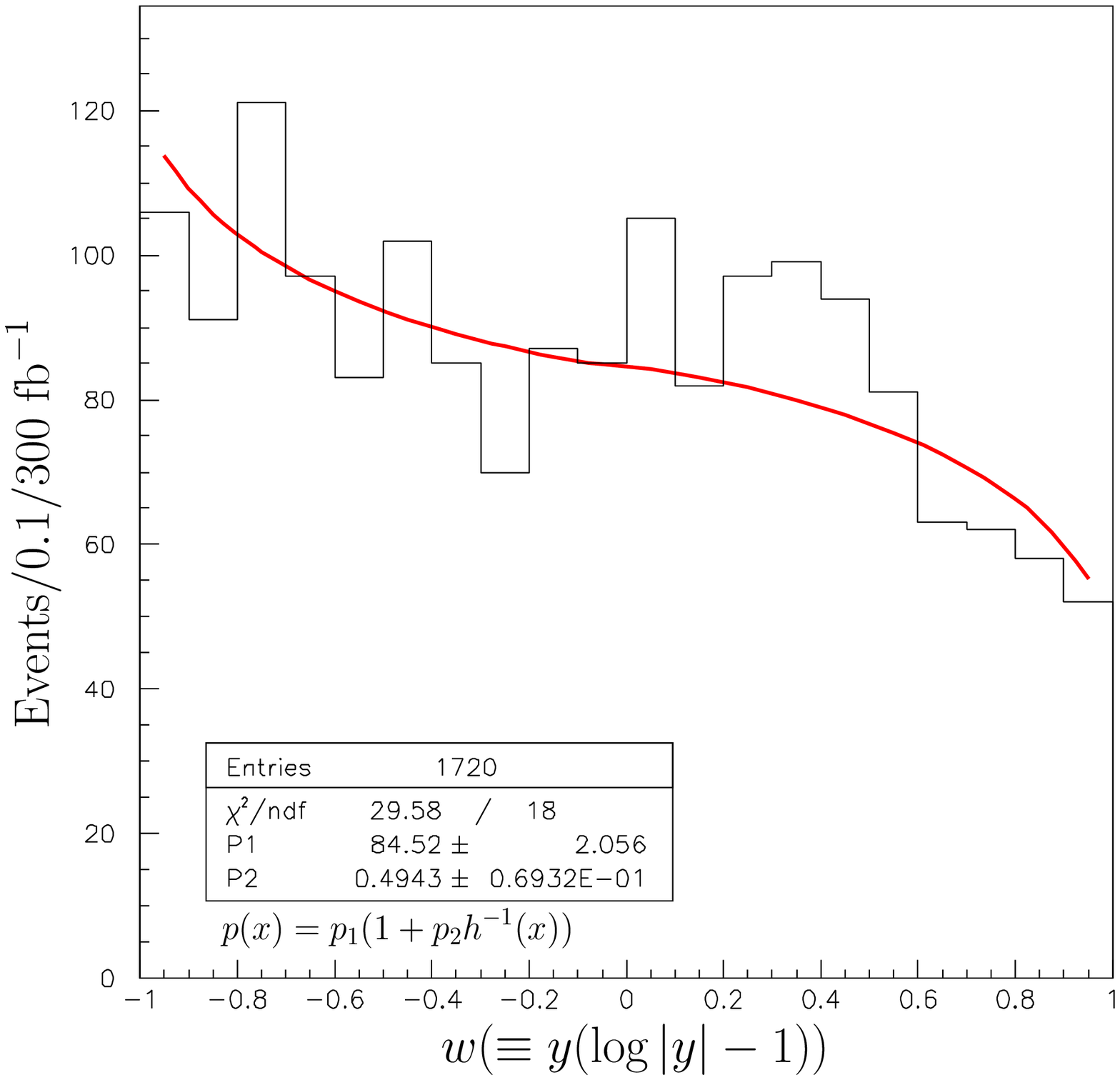}
\end{center}
\caption{The reconstructed $w (\equiv \cos \theta_1 \cos \theta_2 (\log
 | \cos \theta_1 \cos \theta_2 | - 1))$ distribution. A deviation from
 the flat distribution indicates the presence of the spin correlations
 and parity violation in the chargino and the neutralino decays. The
 false solutions are included. A selection cut $\sqrt{\hat s} > 900$~GeV
 is imposed on both of the solutions in each event. The right figure is
 the same but with 300~fb$^{-1}$ of data.}  \label{fig:coscos}
\end{figure}

Although the $\cos \theta_1$ and $\cos \theta_2$ distributions are
trivial in this particular model due to $a_W = 0.0$, there can be a
non-trivial correlation between $\cos \theta_1$ and $\cos \theta_2$.
An example is the distribution of the product $\cos \theta_1 \cos
\theta_2$ which measures the product of the parity asymmetries in the
neutralino and chargino decays independent of $a_W$ (see
Eq.~(\ref{eq:coscos})).

We define a variable,
\begin{eqnarray}
  w = h(y) \equiv y ( \log | y | - 1 ),\ \ \ y = \cos \theta_1 \cos \theta_2.
\end{eqnarray}
The theoretical distribution in Eq.~(\ref{eq:coscos}) in terms of the
variable $w$ is
\begin{eqnarray}
   d\sigma \propto (1 + a_N \langle f_2 \rangle h^{-1}(w) ) d w,
\label{eq:w-distri}
\end{eqnarray}
where $h^{-1}$ is the inverse function of $h(y)$, i.e., $y=h^{-1}(w)$,
and $-1 \leq w \leq 1$. The $w$ distribution is flat in the parity
conserving case ($a_N = 0$). The deviation from the flat distribution is
a signature of parity violation.
The averaged value of $\langle f_2 \rangle$ depends on $\sqrt{\hat
s}_{\rm min}$:
\begin{eqnarray}
 \langle f_2 (600~{\rm GeV}) \rangle = 0.58, \ \ \ 
 \langle f_2 (900~{\rm GeV}) \rangle = 0.75.
\end{eqnarray}
Therefore, with the strategy for the selection cut discussed before, we
expect an asymmetry $a_N \langle f_2 \rangle / 2 \simeq 0.38$ in this
model, where the factor of two is the effect of fake solutions.

The reconstructed $w$ distribution is shown in Fig.~\ref{fig:coscos}
where we see deviation from the flat distribution. The right figure is
the same analysis with 300~fb$^{-1}$ of data.
We fit the histogram with the function in Eq.~(\ref{eq:w-distri}) and
obtained a significant asymmetry, $a_N \langle f_2 \rangle / 2 \simeq
0.47 \pm 0.12$ ($0.49 \pm 0.07$) which deviates from zero by 4$\sigma$
(7$\sigma$) with 100~fb$^{-1}$ (300~fb$^{-1}$) of data. A somewhat
larger value compared to the expectation (0.38) can be understood by the
fact that the effective $\sqrt{\hat s}_{\rm min}$ is larger than 900~GeV
because we have imposed a cut on both of the solutions.

Observation of this distribution together with the measurement of $a_N$
by the $z_l$ distribution will be a quite interesting confirmation of
the spin-spin correlations.

\subsubsection{\boldmath $\phi_1$ distribution}

\begin{figure}[t]
\begin{center}
 \includegraphics[width=7.5cm]{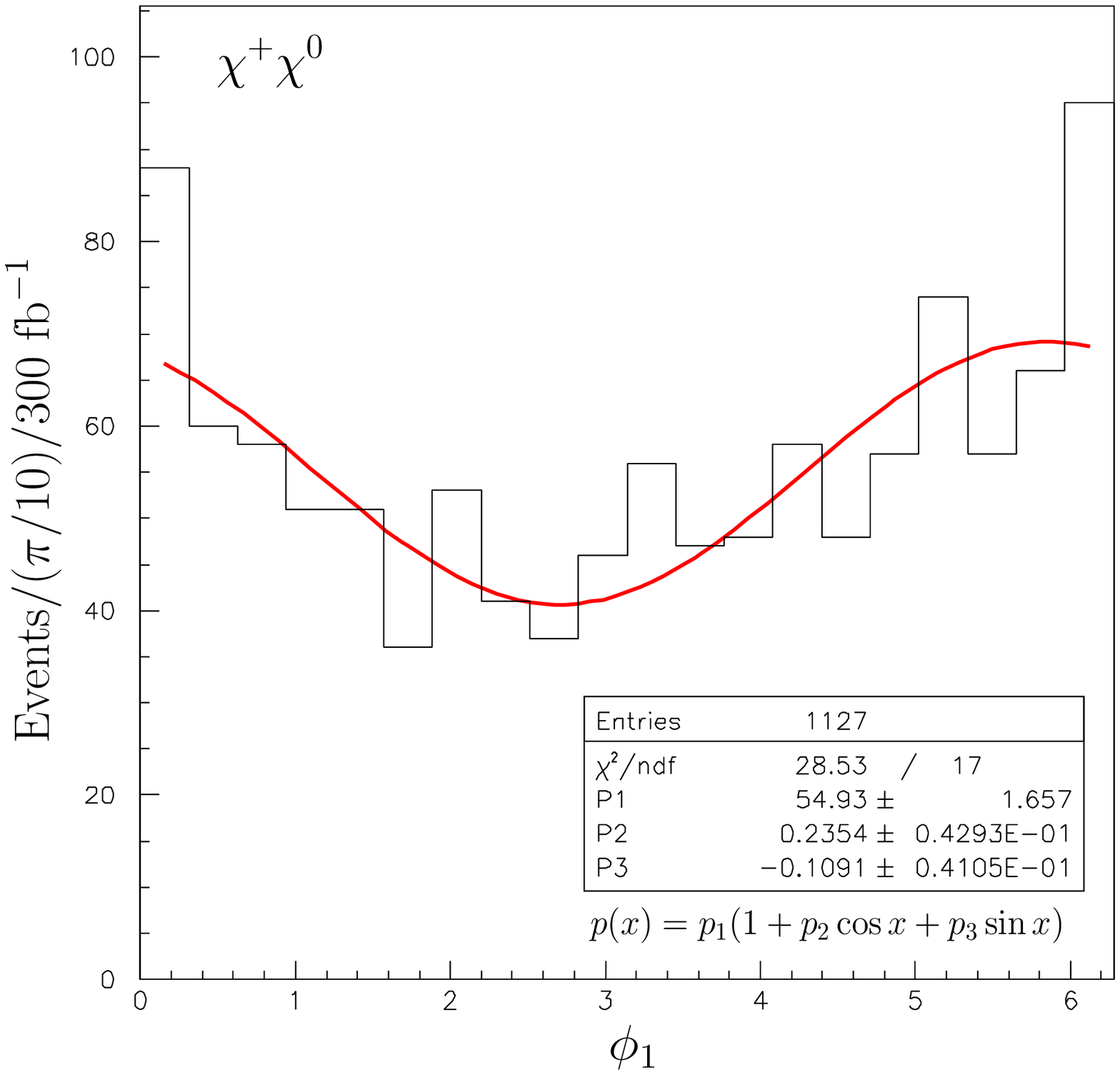}
 \includegraphics[width=7.5cm]{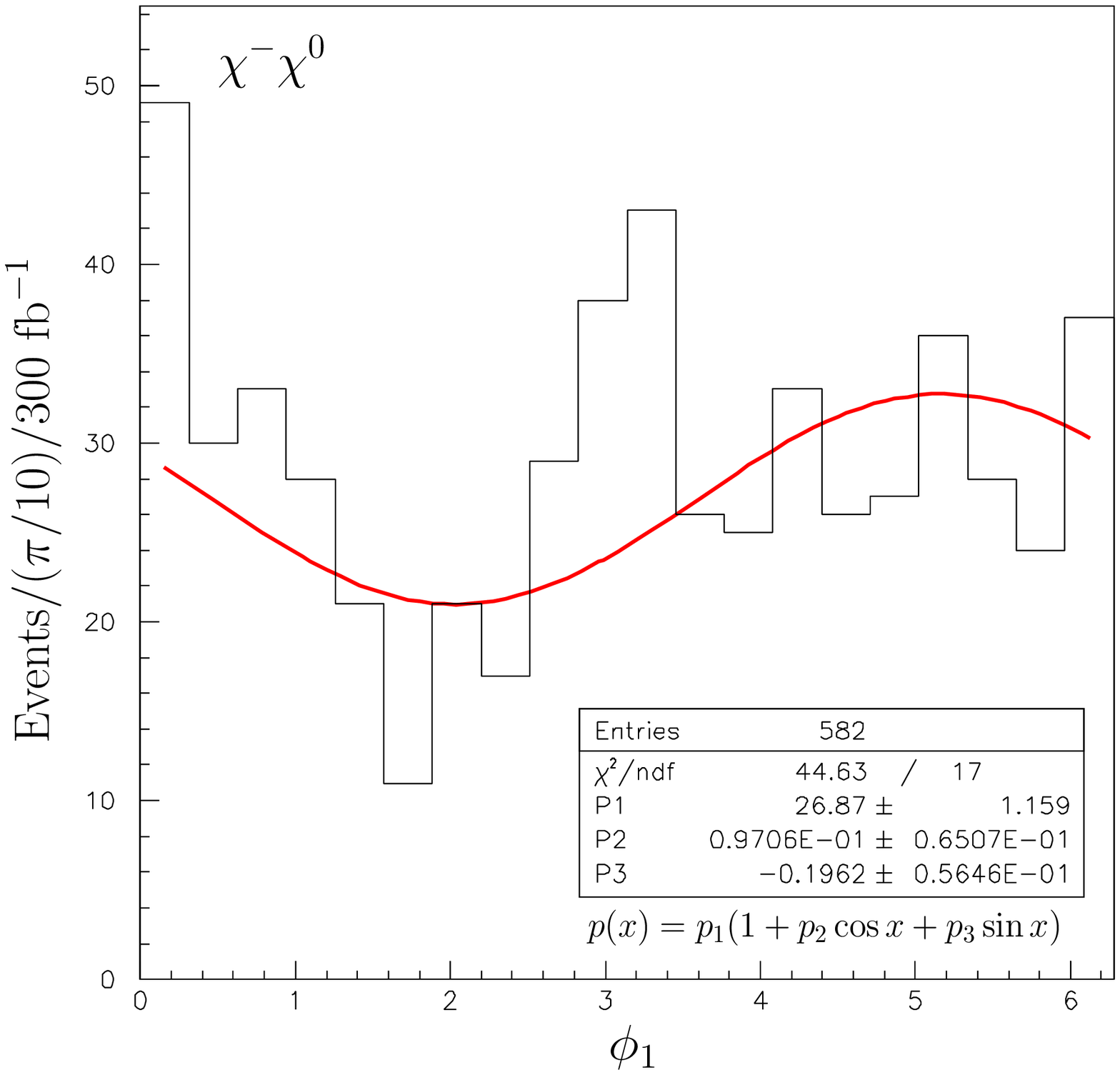}
\end{center}
\caption{The reconstructed $\phi_1$ distributions for the $\chi^+
 \chi^0$ (left) and the $\chi^- \chi^0$ (right) production events.}
 \label{fig:phi1}
\end{figure}

Non-trivial azimuthal-angle distributions show up when there is parity
and/or CP violation in the decay vertex. In order to measure this, we
need to completely reconstruct the kinematics such as the angle
$\theta$. The angle $\phi_1$ is expressed in terms of the angle $\theta$
and the three-momentum of $\tau^\pm$ in the CM frame:
\begin{eqnarray}
 \tan \phi_1^\prime = {
(P^y_{\tilde \tau})_{\rm CM}
\over
(P^x_{\tilde \tau})_{\rm CM} \cos \theta - (P^z_{\tilde \tau})_{\rm
CM} \sin \theta
},\ \ \ \left( 0 \leq \phi_1^\prime \leq \pi \right),
\end{eqnarray}
with
\begin{eqnarray}
 \left \{ 
\begin{array}{ll}
 \phi_1 = \phi_1^\prime  & 
({\rm if}\ (P^y_{\tilde \tau})_{\rm CM} \geq 0 ) \\
 \phi_1 = \phi_1^\prime + \pi & 
({\rm if}\ (P^y_{\tilde \tau})_{\rm CM} < 0)\\
\end{array}
\right.
.
\end{eqnarray}
In order to define the CM frame in Fig.~\ref{fig:coordinate}, we need to
know the direction of the anti-quark which can be determined only
statistically in $pp$-collision experiments.
We take the $z$-direction to be the same direction as that of the total
momentum, $P^z = P_{\chi^\pm}^z + P_{\chi^0}^z$, in the laboratory frame
since the $\bar q$ parton tends to carry a smaller momentum.  In order
to reduce the number of mis-choices, we impose a cut: $P^z > 1200$~GeV.

The $\phi_1$ dependent part of the distribution in Eq.~(\ref{eq:phi1})
has opposite signs for $\chi^+$ and $\chi^-$ productions. We do not
impose a cut on $\hat s$ because the $g_1$ function in Eq.~(\ref{eq:g1})
takes its maximum value at the threshold production. (In order to look
for a CP asymmetry, it may be better to impose a cut. See
Eq.~(\ref{eq:g2}).)
We also use both solutions for $P^z_\nu$. The averaged value of the
functions $g_1$ and $g_2$ are:
\begin{eqnarray}
  {\pi^2 \over 16} \langle
 g_1
\rangle = 0.51,\ \ \ 
  {\pi^2 \over 16} \langle
 g_2
\rangle = 0.16.
\end{eqnarray}
We expect that these values will be affected due to the existence of the
false solution.

The distributions are shown in Fig.~\ref{fig:phi1} with 300~fb$^{-1}$ of
data. The left figure is the distribution of the $\chi^+ \chi^0$
events (i.e., the events with a positive-charge lepton) and the right
figure is from the $\chi^- \chi^0$ events. We fit the histogram by a
function:
\begin{eqnarray}
 p(\phi_1) = p_1 ( 1 + p_2 \cos \phi_1 + p_3 \sin \phi_1 ).
\end{eqnarray}
A qualitatively correct behavior is obtained in the $\chi^+ \chi^0$
events, i.e., $p_2 > 0$ and $p_3 = 0$, but $\chi^- \chi^0$ events do not
show the expected behavior of $p_2 < 0$ and $p_3 = 0$ due to poor
statistics and the selection cut on $P^z$.
We can see from the figures that the selection cut on $P^z$ tends to
give a fake distribution peaked near $\phi_1 \sim 0$ and $2\pi$ for both
$\chi^+ \chi^0$ and $\chi^- \chi^0$ events. One may be able to avoid
this by imposing the $P^z$ cut on both solutions as we have done in the
study of the $\cos \theta_1$ distribution, but it significantly reduces
the statistics. A looser cut on $P^z$ results in a fake distribution by
the mis-choice of the $z$-direction. Nevertheless, it is not a problem
for observing a non-trivial distribution since the theoretic
distribution is different for $\chi^+ \chi^0$ and $\chi^- \chi^0$
productions. For example, one can try to rescale the histogram of the
$\chi^- \chi^0$ events and subtract from (or add to) that of the $\chi^+
\chi^0$ events in order to eliminate (or understand) the fake
distribution.

\subsubsection{\boldmath $\phi_2$ distribution}

\begin{figure}[t]
\begin{center}
 \includegraphics[width=7.5cm]{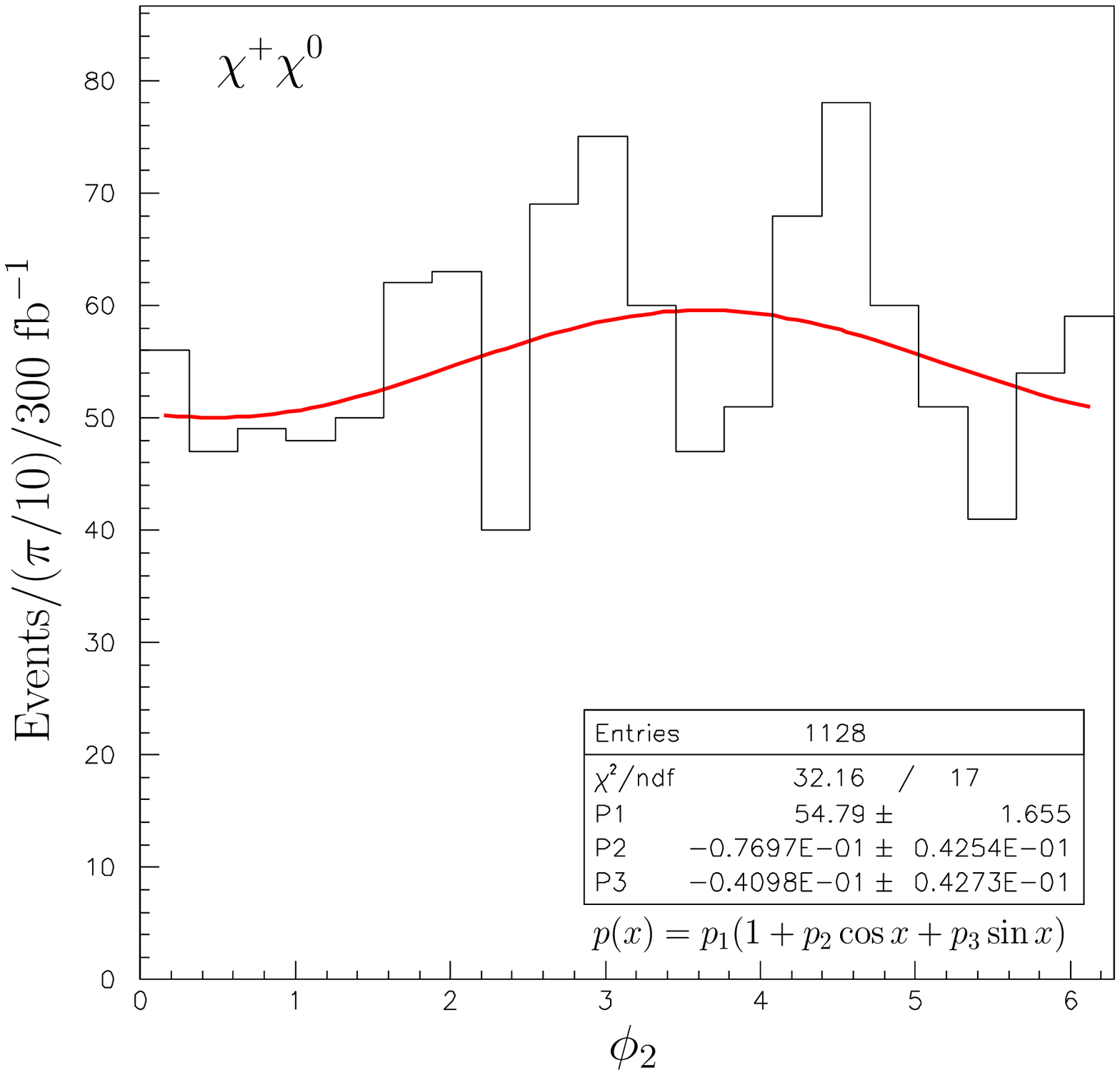}
 \includegraphics[width=7.5cm]{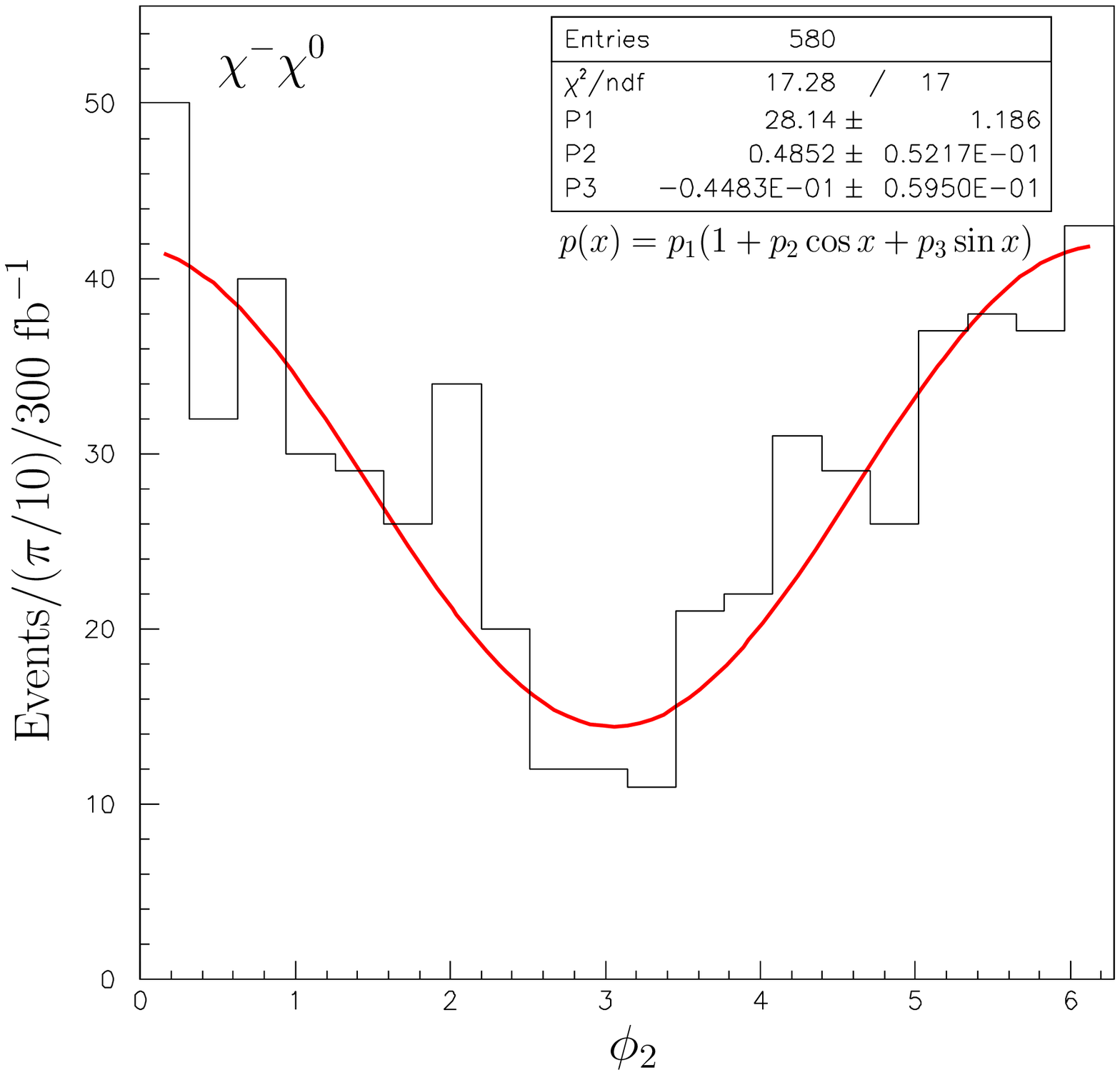}
\end{center}
\caption{The reconstructed $\phi_2$ distributions for the $\chi^+
 \chi^0$ (left) and the $\chi^- \chi^0$ (right) production events.}
 \label{fig:phi2}
\end{figure}

The $\phi_2$ distribution can be obtained by the same method. The
distribution clearly shows an expected behavior in Eq.~(\ref{eq:phi2})
with $a_N = 1.0$ and $\eta_W = 0.0$.

\section{Summary}

If $\tilde \tau$ is the lightest among the superpartners of the Standard
Model particles, the SUSY signatures at the LHC experiments will be very
different from the stable neutralino scenario. We have demonstrated that
the production processes of the neutralinos and charginos have rich
information on model parameters. The spin correlations of intermediate
particles give rise to interesting non-trivial distributions in various
kinematic variables.
In previous studies of SUSY models at hadron colliders, the production
of neutralinos and charginos have been usually thought of as good
processes to discover SUSY through multi-lepton final states. In the
stable neutralino scenario it is nevertheless challenging to extract
information on models out of those processes because of the small cross
sections and difficult kinematics due to escaping neutralinos. Much
attention has been paid to production processes of colored
superparticles and their cascade decays for the measurements of the
model parameters.
However, as we have shown, the chargino-neutralino production process
may provide us with the best opportunity for understanding SUSY models
in the long-lived $\tilde \tau$ scenario.

The study presented in this paper is not fully realistic in several
senses. We have not included the momentum resolution of the $\tilde
\tau$ tracks or efficiency of the identification. Also, in the study of
various distributions, we have ignored errors in the measurements of the
chargino and the neutralino masses. We have used the transverse missing
momentum evaluated by the fast detector simulator, but the resolution
may be very different in real experiments. The trigger efficiency of the
process has also been ignored.  A more detailed analysis is necessary
when we discover the long-lived $\tilde \tau$.  The analytical formulae
presented in this paper will be useful in such future studies.

This work is inspired by studies of the electroweak theory in
Refs.~\cite{Gaemers:1978hg} where the differential cross section of the
process $e^+ e^- \to W^+ W^-$ is calculated including the effect of spin
correlations. These various distributions are studied for the purpose of
confirming the SU(2)$_L$ $\times$ U(1)$_Y$ gauge interactions. The
density matrix calculated there has been used to put constraints on
anomalous interactions among gauge bosons at the LEP-II
experiments~\cite{Abreu:2001rpa}.
If $\tilde \tau$ is long-lived, the cross-section formula calculated in
this paper can be used as a good test of SUSY at the LHC
experiments just like we have confirmed the Standard Model at the LEP
experiments.

We here comment on the $\chi^0 \chi^0$ production processes which we did
not study in this paper. In many cases, these processes have smaller
cross sections than the $\chi^\pm \chi^0$ process.  Since the $Z$-boson
vertex involving the same mass eigenstates, $Z - \chi^0_i - \chi^0_i$,
vanishes identically, the main production process is $\chi^0_i \chi^0_j$
with $i \neq j$. If we require the opposite charges for two $\tilde
\tau$'s and the leptonic decays for both of the $\tau$ leptons, the
number of events will get smaller. However, for the reconstruction
of the kinematics, it is much simpler than the $\chi^\pm \chi^0$
production events. We can reconstruct the final state without the
knowledge of the neutralino masses. Moreover, there is no discrete
ambiguity for the reconstruction. The study of these processes will also
be important if $\tilde \tau$ is long-lived, although it may be
challenging due to the limited statistics.

\section*{Acknowledgments}

I thank Michael Graesser for reading the manuscript, stimulating
discussions and useful comments. I also thank Alex Friedland for
discussions and useful comments.


\begin{thebibliography}{0}
\bibitem{Gherghetta:1998tq}
  T.~Gherghetta, G.~F.~Giudice and A.~Riotto,
  Phys.\ Lett.\  B {\bf 446}, 28 (1999)
  [arXiv:hep-ph/9808401];
  T.~Asaka, K.~Hamaguchi and K.~Suzuki,
  Phys.\ Lett.\  B {\bf 490}, 136 (2000)
  [arXiv:hep-ph/0005136];
  J.~L.~Feng, S.~Su and F.~Takayama,
  Phys.\ Rev.\  D {\bf 70}, 075019 (2004)
  [arXiv:hep-ph/0404231];
  M.~Pospelov,
  Phys.\ Rev.\ Lett.\  {\bf 98}, 231301 (2007)
  [arXiv:hep-ph/0605215];
  K.~Kohri and F.~Takayama,
  Phys.\ Rev.\  D {\bf 76}, 063507 (2007)
  [arXiv:hep-ph/0605243];
  M.~Kawasaki, K.~Kohri and T.~Moroi,
  Phys.\ Lett.\  B {\bf 649}, 436 (2007)
  [arXiv:hep-ph/0703122];
  J.~Pradler and F.~D.~Steffen,
  arXiv:0710.2213 [hep-ph];
  K.~Jedamzik,
  JCAP {\bf 0803}, 008 (2008)
  [arXiv:0710.5153 [hep-ph]];
  M.~Kawasaki, K.~Kohri, T.~Moroi and A.~Yotsuyanagi,
  arXiv:0804.3745 [hep-ph].

\bibitem{Moroi:1993mb}
  T.~Moroi, H.~Murayama and M.~Yamaguchi,
  Phys.\ Lett.\  B {\bf 303}, 289 (1993);
  E.~Holtmann, M.~Kawasaki, K.~Kohri and T.~Moroi,
  Phys.\ Rev.\  D {\bf 60}, 023506 (1999)
  [arXiv:hep-ph/9805405];
  J.~L.~Feng, A.~Rajaraman and F.~Takayama,
  Phys.\ Rev.\ Lett.\  {\bf 91}, 011302 (2003)
  [arXiv:hep-ph/0302215];
  J.~L.~Feng, A.~Rajaraman and F.~Takayama,
  Phys.\ Rev.\  D {\bf 68}, 063504 (2003)
  [arXiv:hep-ph/0306024];
  J.~L.~Feng, S.~f.~Su and F.~Takayama,
  Phys.\ Rev.\  D {\bf 70}, 063514 (2004)
  [arXiv:hep-ph/0404198];
  M.~Kawasaki, K.~Kohri and T.~Moroi,
  Phys.\ Lett.\  B {\bf 625}, 7 (2005)
  [arXiv:astro-ph/0402490];
  M.~Kawasaki, K.~Kohri and T.~Moroi,
  Phys.\ Rev.\  D {\bf 71}, 083502 (2005)
  [arXiv:astro-ph/0408426].


\bibitem{Ibe:2007km}
  M.~Ibe and R.~Kitano,
  JHEP {\bf 0708}, 016 (2007)
  [arXiv:0705.3686 [hep-ph]].

\bibitem{Hinchliffe:1998ys}
  I.~Hinchliffe and F.~E.~Paige,
  Phys.\ Rev.\  D {\bf 60}, 095002 (1999)
  [arXiv:hep-ph/9812233].

\bibitem{Ellis:2006vu}
  J.~R.~Ellis, A.~R.~Raklev and O.~K.~Oye,
  JHEP {\bf 0610}, 061 (2006)
  [arXiv:hep-ph/0607261].

\bibitem{Nisati:1997gb}
  A.~Nisati, S.~Petrarca and G.~Salvini,
  Mod.\ Phys.\ Lett.\  A {\bf 12}, 2213 (1997)
  [arXiv:hep-ph/9707376].

\bibitem{stau}
G. Polesello and A. Rimoldi, ATLAS Internal Note ATL-MUON-99-006.

\bibitem{Ambrosanio:2000ik}
  S.~Ambrosanio, B.~Mele, S.~Petrarca, G.~Polesello and A.~Rimoldi,
  JHEP {\bf 0101}, 014 (2001)
  [arXiv:hep-ph/0010081].

\bibitem{stauNew}
J.~Ellis, A.~R.~Raklev and O.~K.~Oye, ATLAS Note ATL-PHYS-PUB-2007-016;
	ATL-COM-PHYS-2006-093.

\bibitem{Buchmuller:2004rq}
  W.~Buchmuller, K.~Hamaguchi, M.~Ratz and T.~Yanagida,
  Phys.\ Lett.\  B {\bf 588}, 90 (2004)
  [arXiv:hep-ph/0402179].

\bibitem{Hamaguchi:2004df}
  K.~Hamaguchi, Y.~Kuno, T.~Nakaya and M.~M.~Nojiri,
  Phys.\ Rev.\  D {\bf 70}, 115007 (2004)
  [arXiv:hep-ph/0409248].

\bibitem{Feng:2004yi}
  J.~L.~Feng and B.~T.~Smith,
  Phys.\ Rev.\  D {\bf 71}, 015004 (2005)
  [Erratum-ibid.\  D {\bf 71}, 019904 (2005)]
  [arXiv:hep-ph/0409278].

\bibitem{Rajaraman:2007ae}
  A.~Rajaraman and B.~T.~Smith,
  Phys.\ Rev.\  D {\bf 76}, 115004 (2007)
  [arXiv:0708.3100 [hep-ph]].

\bibitem{Drees:1990yw}
  M.~Drees and X.~Tata,
  Phys.\ Lett.\  B {\bf 252}, 695 (1990).

\bibitem{Feng:1997zr}
  J.~L.~Feng and T.~Moroi,
  Phys.\ Rev.\  D {\bf 58}, 035001 (1998)
  [arXiv:hep-ph/9712499].

\bibitem{Martin:1998vb}
  S.~P.~Martin and J.~D.~Wells,
  Phys.\ Rev.\  D {\bf 59}, 035008 (1999)
  [arXiv:hep-ph/9805289].

\bibitem{Dimopoulos:1996yq}
  S.~Dimopoulos, S.~D.~Thomas and J.~D.~Wells,
  Nucl.\ Phys.\  B {\bf 488}, 39 (1997)
  [arXiv:hep-ph/9609434].

\bibitem{Gupta:2007ui}
  S.~K.~Gupta, B.~Mukhopadhyaya and S.~K.~Rai,
  Phys.\ Rev.\  D {\bf 75}, 075007 (2007)
  [arXiv:hep-ph/0701063].

\bibitem{Barger:1983wc}
  V.~D.~Barger, R.~W.~Robinett, W.~Y.~Keung and R.~J.~N.~Phillips,
  Phys.\ Lett.\  B {\bf 131}, 372 (1983).

\bibitem{Haber:1994pe}
  H.~E.~Haber,
  arXiv:hep-ph/9405376.

\bibitem{Bullock:1992yt}
  B.~K.~Bullock, K.~Hagiwara and A.~D.~Martin,
  Nucl.\ Phys.\  B {\bf 395}, 499 (1993).

\bibitem{Buckley:2007th}
  M.~R.~Buckley, H.~Murayama, W.~Klemm and V.~Rentala,
  arXiv:0711.0364 [hep-ph].

\bibitem{Corcella:2002jc}
  G.~Corcella {\it et al.},
  arXiv:hep-ph/0210213.

\bibitem{Richardson:2001df}
  P.~Richardson,
  JHEP {\bf 0111}, 029 (2001)
  [arXiv:hep-ph/0110108].

\bibitem{Moretti:2002eu}
  S.~Moretti, K.~Odagiri, P.~Richardson, M.~H.~Seymour and B.~R.~Webber,
  JHEP {\bf 0204}, 028 (2002)
  [arXiv:hep-ph/0204123].

\bibitem{Lai:1999wy}
  H.~L.~Lai {\it et al.}  [CTEQ Collaboration],
  Eur.\ Phys.\ J.\  C {\bf 12}, 375 (2000)
  [arXiv:hep-ph/9903282].

\bibitem{Jadach:1993hs}
  S.~Jadach, Z.~Was, R.~Decker and J.~H.~Kuhn,
  Comput.\ Phys.\ Commun.\  {\bf 76}, 361 (1993).

\bibitem{RichterWas:2002ch}
  E.~Richter-Was,
  arXiv:hep-ph/0207355.

\bibitem{Lester:1999tx}
  C.~G.~Lester and D.~J.~Summers,
  Phys.\ Lett.\  B {\bf 463}, 99 (1999)
  [arXiv:hep-ph/9906349].

\bibitem{Kawagoe:2004rz}
  K.~Kawagoe, M.~M.~Nojiri and G.~Polesello,
  Phys.\ Rev.\  D {\bf 71}, 035008 (2005)
  [arXiv:hep-ph/0410160].

\bibitem{Cheng:2007xv}
  H.~C.~Cheng, J.~F.~Gunion, Z.~Han, G.~Marandella and B.~McElrath,
  JHEP {\bf 0712}, 076 (2007)
  [arXiv:0707.0030 [hep-ph]].

\bibitem{CMS}
M.~Davids {\it et al.}, CMS Note 2006/077.


\bibitem{Gaemers:1978hg}
  K.~J.~F.~Gaemers and G.~J.~Gounaris,
  Z.\ Phys.\  C {\bf 1}, 259 (1979);
  K.~Hagiwara, R.~D.~Peccei, D.~Zeppenfeld and K.~Hikasa,
  Nucl.\ Phys.\  B {\bf 282}, 253 (1987);
  P.~Mery, M.~Perrottet and F.~M.~Renard,
  Z.\ Phys.\  C {\bf 36}, 249 (1987);
  M.~S.~Bilenky, J.~L.~Kneur, F.~M.~Renard and D.~Schildknecht,
  Nucl.\ Phys.\  B {\bf 409}, 22 (1993).

\bibitem{Abreu:2001rpa}
  P.~Abreu {\it et al.}  [DELPHI Collaboration],
  Phys.\ Lett.\  B {\bf 502} (2001) 9
  [arXiv:hep-ex/0102041];
  P.~Achard {\it et al.}  [L3 Collaboration],
  Phys.\ Lett.\  B {\bf 586}, 151 (2004)
  [arXiv:hep-ex/0402036];
  S.~Schael {\it et al.}  [ALEPH Collaboration],
  Phys.\ Lett.\  B {\bf 614} (2005) 7;
  J.~Abdallah {\it et al.}  [DELPHI Collaboration],
  Eur.\ Phys.\ J.\  C {\bf 54}, 345 (2008)
  [arXiv:0801.1235 [hep-ex]].




\end{thebibliography}
\end{document}